\documentclass[twocolumn,showpacs,superscriptaddress,floatfix,prl]{revtex4-1}
\usepackage{graphicx,amsfonts,amssymb,amsmath,hyperref,hypcap,enumerate}

\begin{document}
\title{Topological Exciton Bands in Moir\'e Heterojunctions}

\author{Fengcheng Wu}
\affiliation{Department of Physics, University of Texas at Austin, Austin, TX 78712, USA}
\affiliation{Materials Science Division, Argonne National Laboratory, Argonne, IL 60439, USA}

\author{Timothy Lovorn}
\affiliation{Department of Physics, University of Texas at Austin, Austin, TX 78712, USA}

\author{A. H. MacDonald}
%\email{macdpc@physics.utexas.edu}
\affiliation{Department of Physics, University of Texas at Austin, Austin, TX 78712, USA}

\date{\today}

\begin{abstract}
Moir\'e patterns are common in van der Waals heterostructures and can be used to apply
periodic potentials to elementary excitations.
We show that the optical absorption spectrum of transition metal dichalcogenide bilayers is profoundly altered by 
long period moir\'e patterns that introduce twist-angle dependent satellite excitonic peaks.  
Topological exciton bands with non-zero Chern numbers that support chiral excitonic edge states can be engineered
by combining three ingredients: i) the valley Berry phase induced by electron-hole exchange interactions, 
ii) the moir\'e potential, and iii) the valley Zeeman field.  
\end{abstract}
%\pacs{71.35.-y, 74.78.Fk, 78.67.-n, 03.65.Vf}
% 71.35.-y: Excitons and related phenomena
% 74.78.Fk: Multilayers, superlattices, heterostructures
% 78.67.-n Optical properties of low-dimensional, mesoscopic, and nanoscale materials and structures
% 03.65.Vf Phases: geometric; dynamic or topological

\maketitle

\begin{figure*}[t]
	\includegraphics[width=1.9\columnwidth]{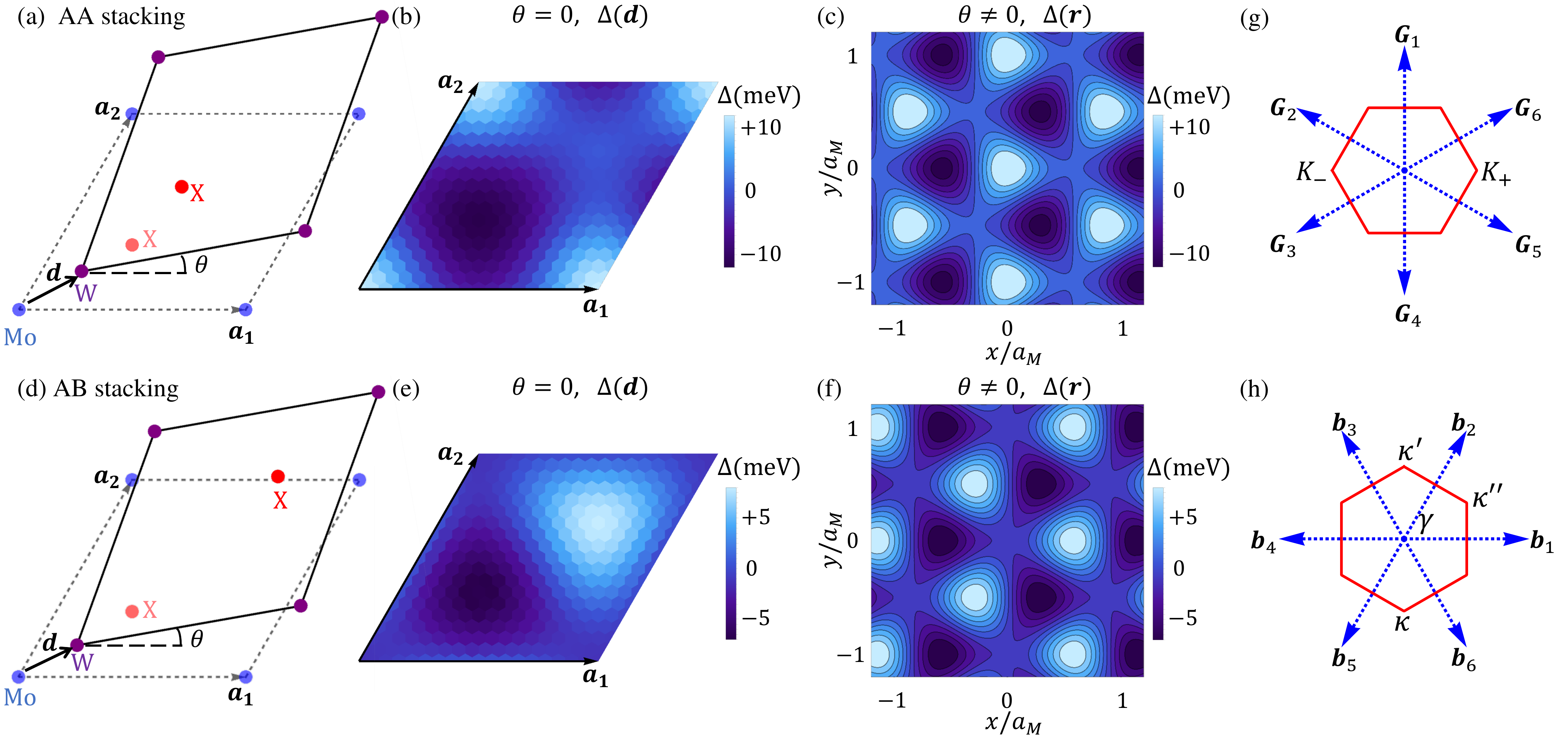}
	\caption{(color online).  (a) Illustration of AA stacking with a small twist angle $\theta$ and an in-plane displacement $\boldsymbol{d}$. 
	(b) Variation of MoS$_2$ band gap as a function of $\boldsymbol{d}$ in AA stacked MoS$_2$/WS$_2$ bilayer with zero twist angle. 
	(c) Variation of MoS$_2$ band gap as a function of position in AA stacked twisted MoS$_2$/WS$_2$.  
 %The spatial scale in this illustrations is the moir\'e period $a_M \sim a_0/\theta$.
(d)-(f) Corresponding plots for AB stacking. (g) First-shell reciprocal lattice vectors $\boldsymbol{G}_j$ of a monolayer TMD triangular lattice and the corresponding Brillouin zone (red hexagon). (h) Moir\'e reciprocal lattice vectors and corresponding Brillouin zone}
	\label{Fig:Illustration}
\end{figure*}

\begin{figure}[t]
	\includegraphics[width=0.9\columnwidth]{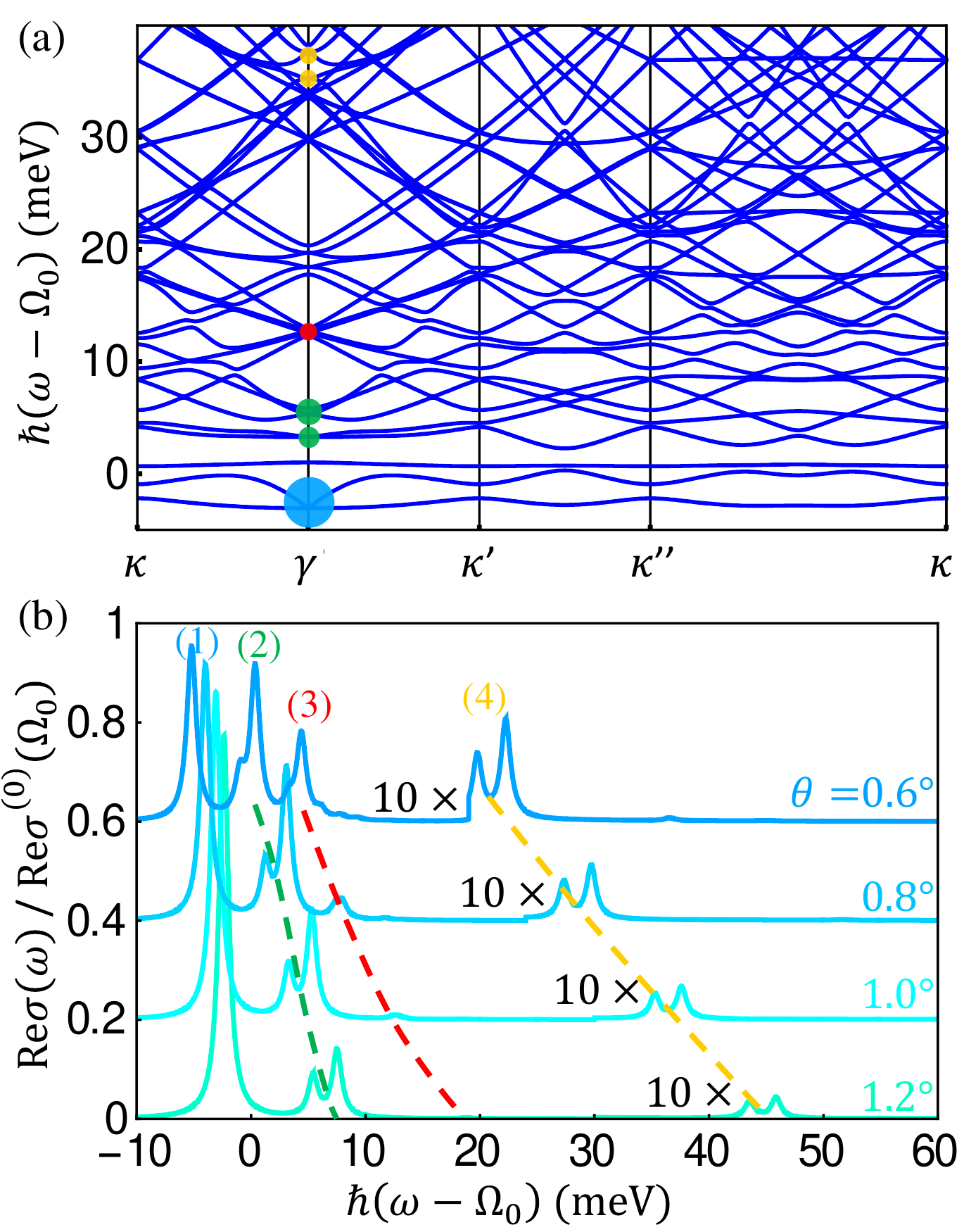}
	\caption{(color online).  (a) Exciton moir\'e bands along paths connecting high symmetry points in the MBZ. The twist angle $\theta$ used 
	for this calculation is $1^{\circ}$. The sizes of the dots at the $\gamma$ point 
	correspond to the weight of state $|K_{\pm}^{(0)}\rangle$. 
        The color of the dots encodes the main character of the states, with blue, red, green and yellow corresponding to 
        character (1), (2), (3) and (4). See text.
(b) Optical conductivity at several twist angles $\theta$. 
Curves for different $\theta$ are shifted vertically for clarity and the dashed curves track the peak evolution. 
The two mini-peaks in (4) are amplified ten times. The broadening factor $\eta$ is taken to be 0.5meV.
The exciton potential parameters of AA stacked MoS$_2$/WS$_2$ were used in these calculations.
	}
	\label{Fig:Band}
\end{figure}

%The properties of the 
Stacking two-dimensional (2D) materials into van der Waals heterostructures opens up new strategies for materials property 
engineering.   One increasingly important example 
is the possibility of using the relative orientation (twist) angle between
two 2D crystals to tune electronic properties.  For small twist angles and lattice-constant mismatches, heterostructures 
exhibit long period moir\'e patterns that can yield dramatic changes.
Moir\'e pattern formed in graphene-based heterostructures has been extensively studied, and many interesting phenomena have been observed,
for example gap opening at graphene's Dirac point \cite{hunt2013, Wang2016}, generation of secondary 
Dirac points\cite{yankowitz2012, ponomarenko2013} and Hofstadter-butterfly spectra in a strong magnetic field\cite{hunt2013, dean2013, Tutuc2016}. 

%In the presence of a strong magnetic field, the moir\'e pattern has enabled clear observations of Hofstadter-butterfly 
%spectra and new fractional quantum Hall states  \cite{hunt2013, dean2013, ponomarenko2013, wang2015}. 

In this Letter, we study the influence of moir\'e patterns on collective excitations, 
focusing on the important case of excitons in the transition metal dichalcogenide (TMD) 2D semiconductors \cite{splendiani2010, mak2010}  
like MoS$_2$ and WS$_2$.  Exciton features dominate the optical response of these materials 
because electron-hole pairs are strongly bound by the Coulomb interaction \cite{ye2014, He2014, chernikov2014, Qiu2013}. 
An exciton inherits a pseudospin-$1/2$ valley degree of freedom from its constituent electron and hole, and 
the exciton valley pseudospin can be optically addressed \cite{cao2012valley, zeng2012valley, mak2012control, Di2012}, 
providing access to the valley Hall effect \cite{mak2014valley} and the valley selective optical Stark effect \cite{kim2014, sie2015}.  

As in the case of graphene/hexagonal-boron-nitride and graphene/graphene bilayers,
a moir\'e pattern can be established in TMD bilayers by using two different materials with 
a small lattice mismatch, by applying a small twist, or by combining both effects. 
TMD heterostructures have been realized \cite{Fang2014,gong2014, liu2014evolution} experimentally and can host interesting effects, for example the observation of valley polarized interlayer excitons with
long lifetimes \cite{rivera2016valley}, the theoretical prediction of multiple degenerate interlayer excitons \cite{Yu2015}, 
and the possibility of achieving spatially indirect exciton condensation \cite{fogler2014high, Wu2015}.
Our focus here is instead on the {\em intralayer} excitons that are more strongly coupled to light.   
As we explain below, the moir\'e pattern produces a periodic potential, mixing momentum
%We approximate the superlattice potential by the spatial variation of the intralayer band gap, which is extracted from {\it ab initio} calculation. This %approximation is suitable because the periodicity of moir\'e pattern at small twist angle is large compared to both lattice constant and the Bohr radius %of exciton.
states separated by moir\'e reciprocal lattice vectors and 
producing satellite optical absorption peaks that are revealing.  
%For example, when a monolayer TMD is placed on a substrate with which it interacts weakly its 
The exciton energy-momentum dispersion can be measured by tracking the dependence of satellite peak energies on 
twist angle.

The valley pseudospin of an exciton is intrinsically coupled to its center-of-mass motion by the electron-hole exchange interaction\cite{yu2014dirac, Glazov2014, Yu2014, wu2015Exciton}.
This effective spin-orbit coupling endows the exciton  with a 
$2\pi$ momentum-space Berry phase\footnote{The effect of the exchange interaction on spatially indirect excitons has been discussed, for example, in Ref.~\cite{Durnev2016}}. 
%The exchange interactions require a finite overlap between electron and hole wave functions and are 
%therefore much weaker for interlayer excitons.
We show that topological exciton bands characterized by quantized 
Chern numbers can be achieved by exploiting this momentum-space Berry phase 
combined with a periodic potential due to the moir\'e pattern 
and time-reversal symmetry breaking by a Zeeman field. 
All three ingredients are readily available in TMD bilayers. 
The bulk topological bands lead to chiral edge states, which can support unidirectional transport of excitons 
optically generated on the edge. Our study therefore suggests a practical new route to engineer topological
collective excitations which are now actively sought\cite{yuen2014topologically, Karzig2015, Bardyn2015, Nalitov2015, Song2016, jin2016}
in several different contexts.

\noindent
{\em Exciton Potential Energy}--- 
For definiteness we consider the common chalcogen 
TMD bilayer MoX$_2$/WX$_2$ with a small twist angle $\theta$ and an in-plane displacement $\boldsymbol{d}$. 
TMDs with a common chalcogen (X) atom have small lattice mismatches ($\sim 0.1$\%), which we neglect to simplify calculations.
Because of the van der Waals heterojunction character and relative band offsets, both conduction and 
valence bands of MoX$_2$ and WX$_2$ are weakly coupled across the heterojunction. 
%This justifies treating the effect of interlayer coupling as a perturbation to intralayer excitons. 
The heterojunctions have two distinct stacking orders AA and AB, which are
illustrated in Figs.~\ref{Fig:Illustration}(a) and \ref{Fig:Illustration}(d).
Both configurations have been experimentally realized \cite{rivera2016valley, wilson2016band}.

We start by analyzing the bilayer electronic structure at zero twist angle. Fully-relativistic density-functional-theory {\em ab initio} calculation is performed for crystalline MoS$_2$/WS$_2$ ($\theta=0^{\circ}$) as a function of relative displacement $\boldsymbol{d}$.
We used the local density approximation with optimized norm-conserving pseudopotentials \cite{hamann2013, schlipf2015}
as implemented in Quantum Espresso \cite{giannozzi2009}, and determined the orbital character of electronic bands using Wannier90 \cite{mostofi2014}.
More details of the calculation are presented in the Supplemental Material \cite{SM}.
Our primary interest here is intra-layer physics.  Figs.~\ref{Fig:Illustration}(b) and \ref{Fig:Illustration}(e) illustrate the 
$\boldsymbol{d}$ dependence of the energy gap at the $K_\pm$ points between states concentrated in the 
MoS$_2$ layer.  
The variation of the gap is a periodic function of $\boldsymbol{d}$ with the 2D lattice periodicity, and is 
adequately approximated by the lowest harmonic expansion:
\begin{equation}
\begin{aligned}
\Delta(\boldsymbol{d})    \equiv E_g(\boldsymbol{d})- \langle {E_g} \rangle
 \approx \sum_{j=1}^6 V_j \exp( i \boldsymbol{G}_j\cdot \boldsymbol{d}),
\end{aligned}
\end{equation}
where $E_g$ is the intralayer band gap of MoS$_2$, $\langle {E_g} \rangle$  is its average over $\boldsymbol{d}$,
and $\boldsymbol{G}_j$ is a one of the first-shell reciprocal lattice vectors illustrated in Fig.~\ref{Fig:Illustration}(g).
Three-fold rotational symmetry of the lattice leads to the constraint:
\begin{equation}
V_1=V_3=V_5,\, V_2=V_4=V_6.
\end{equation}
Because $\Delta$ is real, we also have that $V_1=V_4^*$.  It follows that all six $V_j$ are fixed by $V_1=V \exp(i \psi)$. 
%This parameterization also satisfies the lattice mirror symmetry: $\Delta(d_x, d_y)=\Delta(-d_x, d_y)$.
For MoS$_2$ on WS$_2$ we find that $(V, \psi) = (2.3 {\rm meV}, 30.8^{\circ})$ for AA stacking and $(1.4 {\rm meV}, 98.6^{\circ})$
for AB stacking.  Because the 
%The smaller gap variation in AB structure indicates weaker interlayer coupling, which is likely related to the larger band offset of the spin-like valence %bands in this structure. 
band offset between the two layers can be modified by external electric fields, we expect that the values of
these parameters can be tuned using gate voltages.

Rotation by angle $\theta$ transforms lattice vector $\boldsymbol{L}$  to 
$\boldsymbol{L}' = \mathcal{R}(\theta) \boldsymbol{L}$, where $\mathcal{R}(\theta)$ is the rotation matrix.  
For small twist angles, the relative displacement\footnote{
%Although the translational symmetry of bilayers is generically broken by twists, a new 
%approximate long wavelength periodicity, the moir\'e pattern, is established.  
Bulk properties of moir\'e systems with finite twist angle are independent of the relative displacement prior to 
twist which we set to zero.  See Ref.~\onlinecite{Bistritzer2011} for an explanation. }
%of the  irrelevance of atomic commensurability in semiconductors and semimetals.}
between two layers near position $\boldsymbol{L}'$ is therefore,
\begin{equation}
\boldsymbol{d}(\boldsymbol{L}') = \hat{T}\boldsymbol{L}' = \hat{T}(\boldsymbol{L}'-\boldsymbol{L})\approx \hat{T} (\theta \hat{z}\times\boldsymbol{L}'),
\end{equation}
where the operator $\hat{T}$ reduces a vector to the Wigner-Seitz cell of the triangular lattice labeled by $\boldsymbol{L}$.
In the limit of small $\theta$, the displacement varies smoothly with position.
Because the size of an exciton in TMDs ($\sim 1$nm)  is larger than the lattice constant scale, validating a $k \cdot p$ 
description, but much smaller than moir\'e periods, the influence of the displacement on exciton energy is local \cite{wu2014tunable,Jung2014}.  We find that the variation in the band gap dominates over that of the binding energy\cite{SM}. 
For simplicity, we assume that the variation of exciton energy will follow that of local band gap: 
\begin{equation}
\begin{aligned}
&\Delta(\boldsymbol{r}) \approx \Delta(\boldsymbol{d}(\boldsymbol{r})) 
\approx \sum_{j=1}^6 V_j \exp( i \boldsymbol{G}_j\cdot \boldsymbol{d}(\boldsymbol{r}))\\
&\approx \sum_{j=1}^6 V_j \exp( i \boldsymbol{G}_j\cdot (\theta \hat{z}\times\boldsymbol{r}))
= \sum_{j=1}^6 V_j \exp( i \boldsymbol{b}_j\cdot \boldsymbol{r}).
\end{aligned}
\end{equation}
Here $\Delta(\boldsymbol{r})$ acts as an exciton potential energy, and $\boldsymbol{b}_j = \theta \boldsymbol{G}_j \times \hat{z}$ defines the reciprocal lattice vectors of the moir\'e pattern.
The band gap varies periodically in space due to the moir\'e pattern,  as illustrated in Figs.~\ref{Fig:Illustration}(c) and \ref{Fig:Illustration}(f). The moir\'e periodicity $a_M$ is controlled by the twist angle: $a_M \approx a_0/\theta$, where $a_0$ is the lattice constant of a monolayer TMD.

\noindent
{\em Optical response}.---
%The optical response of monolayer TMDs is dominated by excitonic effect. 
%An exciton is an electron-hole pair bound by  Coulomb interaction. 
We study the $A$ exciton, the lowest-energy bright exciton, in monolayer MoS$_2$. 
%In pristine MoS$_2$, the $A$ exciton at zero center-of-mass momentum consists of doubly degenerate states respectively located in $K_+$ and $K_-$ valleys. %At finite momentum, excitons in the two valleys are coupled by electron-hole exchange interaction, which leads to momentum-dependent valley splitting. 
%The $A$ exciton has a valley degree of freedom.
Its low-energy effective Hamiltonian \cite{yu2014dirac, Glazov2014, Yu2014, wu2015Exciton} is:
\begin{equation}
\label{XKE}
\begin{aligned}
H_0  = \quad & (\hbar\Omega_0+\frac{\hbar^2 \boldsymbol{Q}^2}{2 M})\tau_0 +  J |\boldsymbol{Q}|\tau_0 \\
+ & J |\boldsymbol{Q}| \big[\cos(2 \phi_{\boldsymbol{Q}})\tau_x + \sin(2 \phi_{\boldsymbol{Q}}) \tau_y\big],
\end{aligned}
\end{equation}
where $\boldsymbol{Q}$ is exciton momentum, $\hbar\Omega_0$ is its $\boldsymbol{Q}=0$ energy,
$\hbar^2 \boldsymbol{Q}^2/(2M)$ is its center-of-mass kinetic energy, and $\tau_0$ and $\tau_{x,y}$ are respectively identity matrix and Pauli matrices in valley space. 
%The difference between the electron band gap $E_g$ and $\hbar\Omega_0$ is the exciton binding energy. , and $M$ is the exciton total mass. 
In Eq.~(\ref{XKE}), $\phi_{\boldsymbol{Q}}$ is the orientation angle of the 2D vector $\boldsymbol{Q}$, 
$J |\boldsymbol{Q}|\tau_0$ accounts for intravalley electron-hole exchange interactions, and 
the $\tau_{x,y}$ terms account for their intervalley counterparts\footnote{We used static approximation for the exchange interaction. Retardation effect results in intrinsic energy broadening of exciton states in the light cone. See Ref.~\cite{Glazov2014}.}. 
It follows that the 
%Because of the in-plane pseudo-magnetic field in valley space generated by intervalley exchange interactions, 
$A$ exciton has two energy modes, 
\begin{equation}
E_{\pm}(\boldsymbol{Q}) = \hbar\Omega_0+\frac{\hbar^2 \boldsymbol{Q}^2}{2 M} + J |\boldsymbol{Q}| \pm J |\boldsymbol{Q}|.
\end{equation}
%These two modes are degenerate at $\boldsymbol{Q}=0$, but becomes non-degenerate at finite $|\boldsymbol{Q}|$.
Note that $E_+$ has linear dispersion at small $|\boldsymbol{Q}|$, while the lower mode  $E_-$ is quadratic.
From the {\em ab initio} GW Bethe-Salpeter calculation of
Ref.~\cite{Qiu2015} we obtain $M= 1.3 m_0$ and $J=0.4 \text{eV}\cdot \text{\AA}$, where $m_0$ is the free electron mass. 
%As elaborated in Ref.~\cite{Qiu2015}, there can be $O(|\boldsymbol{Q}|^2)$ corrections to the exchange interactions, which we neglect for simplicity. 
%We turn to the effect of the moir\'e pattern on excitons.  The spatial variation of the band gap acts as a potential landscape for electron-hole pair excitation, i.e. exciton. As we limit our study to small twist angle, the moir\'e pattern has a long wavelength and the potential has a negligible effect on coupling excitons in opposite valleys $K_+$ and $K_-$. 
Since the gap variation is guaranteed by time reversal symmetry to be identical for 
$K_+$ and $K_-$ valleys,
 %Thus, the exciton potential due to moir\'e pattern is a scalar in the valley space. 
%We arrive at the Hamiltonian that describes the $A$ exciton in the moir\'e pattern:
the exciton effective Hamiltonian of twisted bilayers is $H=H_0+ \Delta(\boldsymbol{r})\tau_0$.  

%The moir\'e potential $H_M$ couples two exciton states with momentum difference $\boldsymbol{b}_j$.
We numerically diagonalize the Hamiltonian matrix using a plane-wave expansion; exciton  
momentum reduced to the moir\'e Brillouin zone (MBZ) is a good quantum number.  
%As we are mainly interested in low-energy exciton, a finite matrix at each reduced momentum can be constructed by truncating the unfolded momentum space. 
Fig.~\ref{Fig:Band}(a) illustrates the MoS$_2$ $A$ exciton moir\'e bands for a $1^{\circ}$ twist relative to WS$_2$. 
%The exciton moir\'e bands look similar for slightly larger twist angle, for example $\theta = 1.5^{\circ}$. 
Smaller twist angles imply smaller MBZ dimensions and more moir\'e bands in a given energy window.
For twist angles $\theta>0.5^{\circ}$, the wavelength ($640$nm) of 
light that excites $A$ excitons greatly exceeds the moir\'e periodicity ($<$36nm).
It follows that only excitons close to the MBZ center $\gamma$ are optically active.
The real part of the optical conductivity can be expressed as follows,
\begin{equation}
\begin{aligned}
&\text{Re}\sigma(\omega) =   \frac{1}{\omega \mathcal{A}}\sum_{n} \big |\langle \chi_n  |j_x| G \rangle \big|^2\Gamma_1(\omega-\omega_n)\\
\approx & \frac{\big |\langle K_+^{(0)}  |j_x| G \rangle \big|^2}{\omega \mathcal{A}}\sum_{n}\big|\sum_{\alpha=\pm}\langle \chi_n| K_\alpha^{(0)} \rangle \big|^2\Gamma_1(\omega-\omega_n)\\
\approx & \frac{1}{2}\text{Re}\sigma^{(0)}(\Omega_0)\sum_{n}\big| \sum_{\alpha=\pm}\langle \chi_n| K_\alpha^{(0)} \rangle \big|^2\Gamma_2(\omega-\omega_n),
\end{aligned}
\label{sigmaxx}
\end{equation}
where $\mathcal{A}$ is the system area, $\Gamma_m(\omega-\omega_n)=\eta^m/\big[\hbar^2(\omega-\omega_n)^2+\eta^2\big]$ 
and $\eta$ is a broadening parameter. In Eq.~(\ref{sigmaxx}) $j_x$ is the current operator, $|G\rangle$ is the  neutral
semiconductor ground state, $|\chi_n\rangle$ and $\hbar \omega_n$ are the eigenstates and eigenvalues of the 
exciton moir\'e Hamiltonian at the $\gamma$ point, $|K_\alpha^{(0)}\rangle$ is the valley $K_\alpha$ exciton eigenstate 
at zero twist angle, and $\text{Re}\sigma^{(0)}(\Omega_0)$ is the $A$ exciton 
optical conductivity peak also at zero twist angle.  The assumption
underlying Eq.~(\ref{sigmaxx}) is that only the $|K_\pm^{(0)}\rangle$ component in $|\chi_n\rangle$ contributes to the optical response. 
The final form for $\sigma(\omega)$ emphasizes that the exciton moir\'e potential has the effect of 
redistributing the $A$-exciton peak over a series of closely spaced sub-peaks.  

Theoretical optical conductivities for a series of twist angles are illustrated in Fig.~\ref{Fig:Band}(b).
Peaks labelled (1-4) (see caption) correspond respectively to bare excitons at zero momentum, 
$E_-$ excitons at momentum $\boldsymbol{b}_i$, $E_-$ excitons at momentum $\sqrt{3}\boldsymbol{b}_i\times\hat{z}$,
and $E_+$ excitons at momentum $\boldsymbol{b}_i$.  Without the moir\'e pattern there
would only be one peak centered around frequency $\Omega_0$; 
umklapp scattering off the moir\'e potentials unveils the formerly dark finite-momentum excitonic states.
Both (2) and (4) give rise to two mini-peaks with a small energy splitting. 
As the twist angle decreases, $|\boldsymbol{b}_i|$ is reduced and the satellite peaks shift to lower energy and become stronger;
for $\theta \sim 0.6^{\circ}$ satellite peaks (2) and (3) have strength that is comparable to that of peak (1). 
Although peak (4) is weak, it decays more slowly compared to peak (3) when $\theta$ increases. 
The energy difference between peak (2) and (4) provides a direct measurement of the electron-hole exchange interaction
strength.

\begin{figure}[t]
	\includegraphics[width=0.9\columnwidth]{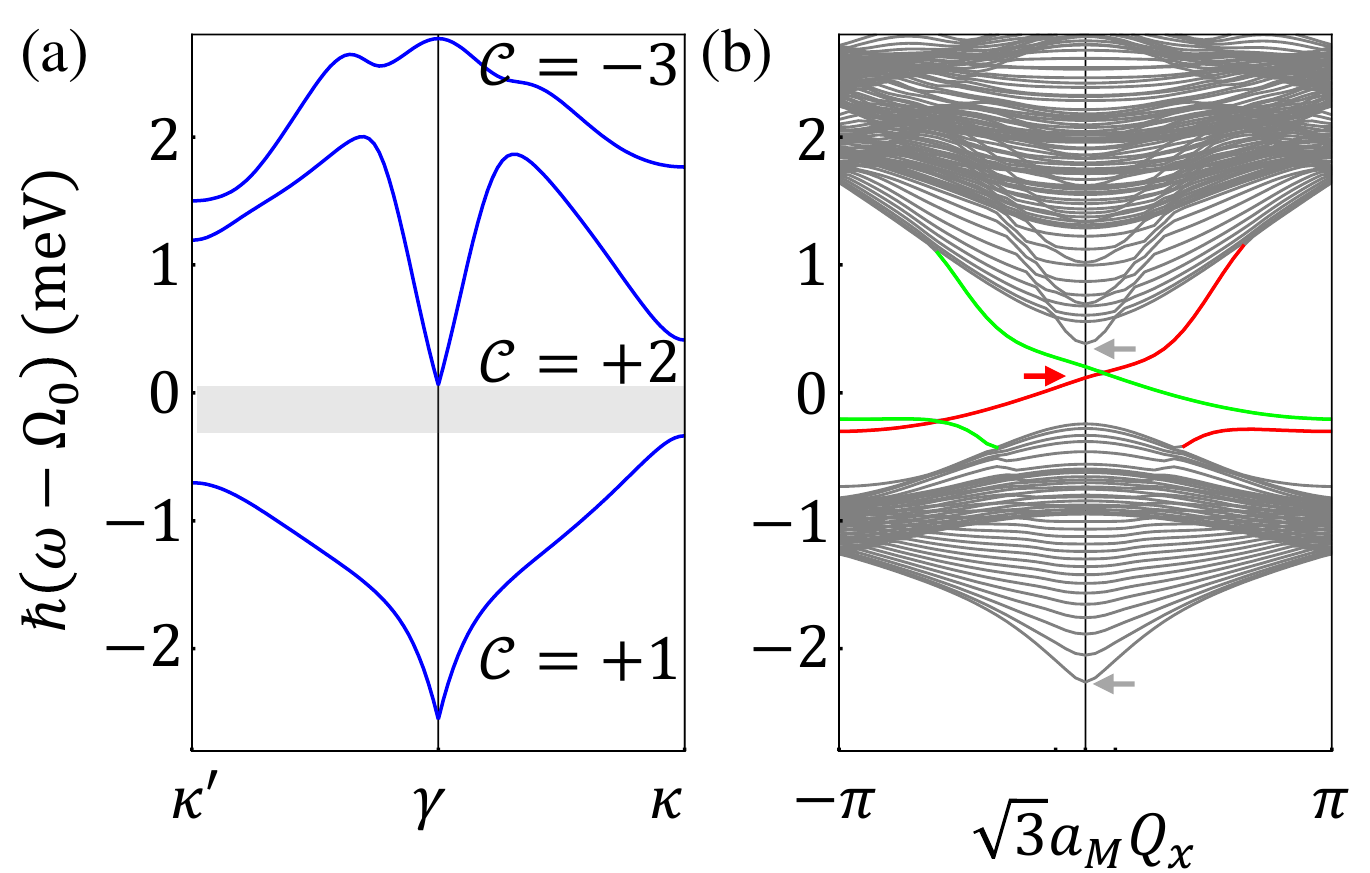}
	\caption{(color online). (a) Topological exciton bands with quantized Chern numbers $\mathcal{C}$ for 
	twist angle $\theta =1^{\circ}$.  The gray bar identifies the gap between the first and second exciton moir\'e band.  
	(b) Stripe geometry quasi-1D bands for the same twist angle, an extended edge along the $x$ direction, and finite width
	 in the $y$ direction. The red and green lines show the dispersions of chiral exciton states that are localized on 
	 opposite edges of the stripe.  
	%(c) Berry curvature $\mathcal{F}$ of the first band in (a). 
%	$\mathcal{F}$ is calculated using the Kubo formula expression.
	 In (a) and (b), $h_z$ is 1.5meV, and ($V$, $\psi$) take the parameter values of AB stacked MoS$_2$/WS$_2$. 
%	(d) Chern number $\mathcal{C}$ of the first exciton moir\'e band as a function of $\theta$ and $\psi$ with ($V$, $h_z$) fixed at
%	 (1.4meV, 1.5meV). The arrows indicate the value of $\psi$ in AA and AB stacked MoS$_2$/WS$_2$. Adding $2\pi/3$ to $\psi$ leads 
%	 to a spatial translation of the moir\'e pattern. It follows that $\mathcal{C}$ is a periodic function of $\psi$ with periodicity $2\pi/3$.
%	 Dashed lines mark approximate phase boundaries obtained by treating moir\'e potential in a perturbation theory\cite{SM}. 
	 }
	\label{Fig:Berry}
\end{figure}

\noindent
{\em Topological excitons}.---The intervalley exchange interaction acts as an in-plane valley-space pseudo-magnetic field
which rotates by 4$\pi$ when the momentum encloses its origin once. 
This non-trivial winding number can be used to engineer topological exciton bands when combined with 
the moir\'e superlattice potential , which provides a finite Brillouin zone and 
an energy gap above the lowest exciton band at every $\boldsymbol{Q}$ point in the MBZ except the $\gamma$ point. 
An external Zeeman term $h_z\tau_z$ can split the degeneracy at $\gamma$.
A Zeeman term of this form has been experimentally realized in monolayer TMDs by applying a 
magnetic field \cite{MacNeil2015, srivastava2015valley, aivazian2015magnetic} and by
using a valley selective optical Stark effect \cite{kim2014, sie2015}. 
The topology of the exciton bands is characterized by Berry curvature $\mathcal{F}$ and 
Chern number $\mathcal{C}$, just as in the electronic case:
\begin{equation}
\begin{aligned}
\mathcal{F}_n(\boldsymbol{Q}) & = \hat{z}\cdot\nabla_{\boldsymbol{Q}}\times\big[ i \langle \chi_n(\boldsymbol{Q}) | \nabla_{\boldsymbol{Q}} | \chi_n(\boldsymbol{Q})\rangle \big],\\
\mathcal{C}_n & = \int_{\text{MBZ}} \frac{d^2\boldsymbol{Q}}{2\pi} \mathcal{F}_n(\boldsymbol{Q}),
\end{aligned}
\end{equation} 
where $|\chi_n(\boldsymbol{Q})\rangle$ represents the $n$th eigenstate of Hamiltonian $H$ at momentum $\boldsymbol{Q}$.
Fig.~\ref{Fig:Berry}(a) presents our results for the topological properties of 
moir\'e exciton bands in AB stacked MoS$_2$/WS$_2$. 
%These results were obtained for Zeeman energy $h_z= 1.5$ meV. 
We find that the first exciton band can possess a non-zero Chern number, 
and that it is isolated from other bands by a global energy gap.
%Interestingly, the second and third bands respectively carry Chern number +2 and -3, although there is no overall energy gap between them. 
The corresponding Berry curvature $\mathcal{F}$ has hot spots around $\gamma$, $\kappa$, and $\kappa'$ points in the MBZ. 
% as illustrated in Fig.~%\ref{Fig:Berry}(c). 
$\mathcal{F}$ around $\gamma$ is simply understood 
in terms of the valley Berry phase induced by the exchange interaction, and its sign is determined by that of $h_z$. 
The peak in $\mathcal{F}$ around the $\kappa$ and $\kappa'$ point is related to gap opening due to moir\'e pattern, 
and can vary as a function of $\psi$, the phase of $V_1$. 
%Fig.~\ref{Fig:Berry}(d) shows the Chern number of the first band as a function of $\psi$ and twist angle $\theta$ with ($V$, $h_z$) fixed to (1.4meV, %1.5meV). Along the phase boundary lines at which the Chern number changes, 
%the energy gaps at $\kappa$ or $\kappa'$ close.  
We find that the Chern number is finite in a large parameter space of $(\psi, \theta)$\cite{SM}.
Therefore we expect that topological exciton bands appear routinely in TMD bilayers.
%\begin{equation}
%\mathcal{F}(\boldsymbol{Q}) \approx \frac{J^2 h_z}{ (h_z^2+J^2|\boldsymbol{Q}|^2)^{3/2} } 
%\end{equation}

\begin{figure}[t]
	\includegraphics[width=1\columnwidth]{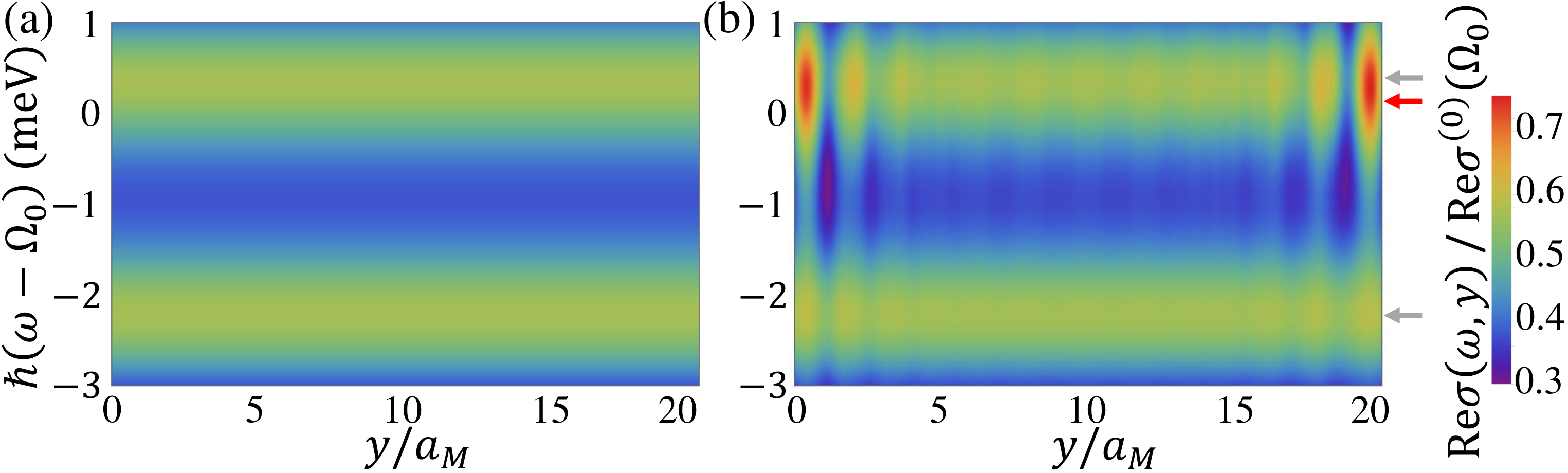}
	\caption{(color online). Spatially resolved optical conductivity on the same stripe geometry as studied in Fig.~\ref{Fig:Berry}(b). The optical conductivity is averaged over $x$. (a) Response from bulk states. The energy splitting between the two peaks is due to the Zeeman energy. An energy broadening factor of 1meV is used. (b) Response from both bulk and edge states. There is an enhancement of the optical response around the edge due to edge states. The arrows indicate the energy level of bulk states (gray arrow) and edge state (red arrow) in the absence of energy broadening, and correspond to the three arrows shown in Fig.~\ref{Fig:Berry}(b).
	 }
	\label{local_response}
\end{figure}

Chiral excitonic edge states expected for 
topological bands are confirmed by studying the energy spectrum of a finite-width stripe \cite{SM}, 
as illustrated in Fig.~\ref{Fig:Berry}(b).  These states can support unidirectional excitonic transport channels. 
We have computed optical response of the edge states\cite{SM}.
The spatially resolved optical conductivity is shown in Fig.~\ref{local_response}.
Based on numerical results, we find that the {\it maximum} local optical conductivity due to one edge state is about $0.19\text{Re}\sigma^{(0)}(\Omega_0)$, which is comparable in magnitude to that of the bulk states.
As illustrated in Fig.~\ref{local_response}(b), edge states give rise to enhanced response around the edge, and therefore can  be detected by spatially resolved absorption spectroscopy.
%It can be identified experimentally by applying spatially resolved photoluminescence or absorption spectroscopy
%to a TMD bilayer which is pumped and probed at an in-gap frequency by a laser spot localized near one edge.  

%As an exciton is an energy carrier, the chiral channel allows energy flow without backscattering.  
In summary, intralayer excitons in a twisted TMD bilayer exhibit rich phenomena enabled by the moir\'e pattern,
including satellite excitonic peaks in optical absorption peaks that are tunable by varying twist angle.
The moir\'e superlattice potential, the exciton Zeeman field,
and the electron-hole exchange induced valley Berry phase can in 
combination give rise to topological exciton bands.
Our analysis points to a practical strategy to realize topological excitons.

We would like to thank Feng Wang and Ivar Martin for useful discussions. 
Work at Austin was supported by the Department of Energy, Office of Basic Energy Sciences under contract DE-FG02-ER45118 and  award \# DE-SC0012670, and by the Welch foundation under grant TBF1473. 
Work of FW at Argonne was supported by the Department of Energy, Office of Science, Materials Sciences and Engineering Division. 
We acknowledge computer time allocations from Texas Advanced Computing Center.

\bibliographystyle{apsrev4-1}
\bibliography{refs}

\begin{thebibliography}{57}%
\makeatletter
\providecommand \@ifxundefined [1]{%
 \@ifx{#1\undefined}
}%
\providecommand \@ifnum [1]{%
 \ifnum #1\expandafter \@firstoftwo
 \else \expandafter \@secondoftwo
 \fi
}%
\providecommand \@ifx [1]{%
 \ifx #1\expandafter \@firstoftwo
 \else \expandafter \@secondoftwo
 \fi
}%
\providecommand \natexlab [1]{#1}%
\providecommand \enquote  [1]{``#1''}%
\providecommand \bibnamefont  [1]{#1}%
\providecommand \bibfnamefont [1]{#1}%
\providecommand \citenamefont [1]{#1}%
\providecommand \href@noop [0]{\@secondoftwo}%
\providecommand \href [0]{\begingroup \@sanitize@url \@href}%
\providecommand \@href[1]{\@@startlink{#1}\@@href}%
\providecommand \@@href[1]{\endgroup#1\@@endlink}%
\providecommand \@sanitize@url [0]{\catcode `\\12\catcode `\$12\catcode
  `\&12\catcode `\#12\catcode `\^12\catcode `\_12\catcode `\%12\relax}%
\providecommand \@@startlink[1]{}%
\providecommand \@@endlink[0]{}%
\providecommand \url  [0]{\begingroup\@sanitize@url \@url }%
\providecommand \@url [1]{\endgroup\@href {#1}{\urlprefix }}%
\providecommand \urlprefix  [0]{URL }%
\providecommand \Eprint [0]{\href }%
\providecommand \doibase [0]{http://dx.doi.org/}%
\providecommand \selectlanguage [0]{\@gobble}%
\providecommand \bibinfo  [0]{\@secondoftwo}%
\providecommand \bibfield  [0]{\@secondoftwo}%
\providecommand \translation [1]{[#1]}%
\providecommand \BibitemOpen [0]{}%
\providecommand \bibitemStop [0]{}%
\providecommand \bibitemNoStop [0]{.\EOS\space}%
\providecommand \EOS [0]{\spacefactor3000\relax}%
\providecommand \BibitemShut  [1]{\csname bibitem#1\endcsname}%
\let\auto@bib@innerbib\@empty
%</preamble>
\bibitem [{\citenamefont {Hunt}\ \emph {et~al.}(2013)\citenamefont {Hunt} \emph
  {et~al.}}]{hunt2013}%
  \BibitemOpen
  \bibfield  {author} {\bibinfo {author} {\bibfnamefont {B.}~\bibnamefont
  {Hunt}} \emph {et~al.},\ }\href
  {http://science.sciencemag.org/content/340/6139/1427} {\bibfield  {journal}
  {\bibinfo  {journal} {Science}\ }\textbf {\bibinfo {volume} {340}},\ \bibinfo
  {pages} {1427} (\bibinfo {year} {2013})}\BibitemShut {NoStop}%
\bibitem [{\citenamefont {Wang}\ \emph {et~al.}(2016)\citenamefont {Wang} \emph
  {et~al.}}]{Wang2016}%
  \BibitemOpen
  \bibfield  {author} {\bibinfo {author} {\bibfnamefont {E.}~\bibnamefont
  {Wang}} \emph {et~al.},\ }\href
  {http://www.nature.com/nphys/journal/v12/n12/full/nphys3856.html} {\bibfield
  {journal} {\bibinfo  {journal} {Nat. Phys.}\ }\textbf {\bibinfo {volume}
  {12}},\ \bibinfo {pages} {1111} (\bibinfo {year} {2016})}\BibitemShut
  {NoStop}%
\bibitem [{\citenamefont {Yankowitz}\ \emph {et~al.}(2012)\citenamefont
  {Yankowitz}, \citenamefont {Xue}, \citenamefont {Cormode}, \citenamefont
  {Sanchez-Yamagishi}, \citenamefont {Watanabe}, \citenamefont {Taniguchi},
  \citenamefont {Jarillo-Herrero}, \citenamefont {Jacquod},\ and\ \citenamefont
  {LeRoy}}]{yankowitz2012}%
  \BibitemOpen
  \bibfield  {author} {\bibinfo {author} {\bibfnamefont {M.}~\bibnamefont
  {Yankowitz}}, \bibinfo {author} {\bibfnamefont {J.}~\bibnamefont {Xue}},
  \bibinfo {author} {\bibfnamefont {D.}~\bibnamefont {Cormode}}, \bibinfo
  {author} {\bibfnamefont {J.~D.}\ \bibnamefont {Sanchez-Yamagishi}}, \bibinfo
  {author} {\bibfnamefont {K.}~\bibnamefont {Watanabe}}, \bibinfo {author}
  {\bibfnamefont {T.}~\bibnamefont {Taniguchi}}, \bibinfo {author}
  {\bibfnamefont {P.}~\bibnamefont {Jarillo-Herrero}}, \bibinfo {author}
  {\bibfnamefont {P.}~\bibnamefont {Jacquod}}, \ and\ \bibinfo {author}
  {\bibfnamefont {B.~J.}\ \bibnamefont {LeRoy}},\ }\href
  {http://www.nature.com/nphys/journal/v8/n5/full/nphys2272.html} {\bibfield
  {journal} {\bibinfo  {journal} {Nat. Phys.}\ }\textbf {\bibinfo {volume}
  {8}},\ \bibinfo {pages} {382} (\bibinfo {year} {2012})}\BibitemShut {NoStop}%
\bibitem [{\citenamefont {Ponomarenko}\ \emph {et~al.}(2013)\citenamefont
  {Ponomarenko} \emph {et~al.}}]{ponomarenko2013}%
  \BibitemOpen
  \bibfield  {author} {\bibinfo {author} {\bibfnamefont {L.}~\bibnamefont
  {Ponomarenko}} \emph {et~al.},\ }\href
  {http://www.nature.com/nature/journal/v497/n7451/full/nature12187.html}
  {\bibfield  {journal} {\bibinfo  {journal} {Nature}\ }\textbf {\bibinfo
  {volume} {497}},\ \bibinfo {pages} {594} (\bibinfo {year}
  {2013})}\BibitemShut {NoStop}%
\bibitem [{\citenamefont {Dean}\ \emph {et~al.}(2013)\citenamefont {Dean} \emph
  {et~al.}}]{dean2013}%
  \BibitemOpen
  \bibfield  {author} {\bibinfo {author} {\bibfnamefont {C.}~\bibnamefont
  {Dean}} \emph {et~al.},\ }\href
  {http://www.nature.com/nature/journal/v497/n7451/full/nature12186.html}
  {\bibfield  {journal} {\bibinfo  {journal} {Nature}\ }\textbf {\bibinfo
  {volume} {497}},\ \bibinfo {pages} {598} (\bibinfo {year}
  {2013})}\BibitemShut {NoStop}%
\bibitem [{\citenamefont {Kim}\ \emph {et~al.}()\citenamefont {Kim},
  \citenamefont {DaSilva}, \citenamefont {Huang}, \citenamefont {Fallahazad},
  \citenamefont {Larentis}, \citenamefont {Taniguchi}, \citenamefont
  {Watanabe}, \citenamefont {LeRoy}, \citenamefont {MacDonald},\ and\
  \citenamefont {Tutuc}}]{Tutuc2016}%
  \BibitemOpen
  \bibfield  {author} {\bibinfo {author} {\bibfnamefont {K.}~\bibnamefont
  {Kim}}, \bibinfo {author} {\bibfnamefont {A.}~\bibnamefont {DaSilva}},
  \bibinfo {author} {\bibfnamefont {S.}~\bibnamefont {Huang}}, \bibinfo
  {author} {\bibfnamefont {B.}~\bibnamefont {Fallahazad}}, \bibinfo {author}
  {\bibfnamefont {S.}~\bibnamefont {Larentis}}, \bibinfo {author}
  {\bibfnamefont {T.}~\bibnamefont {Taniguchi}}, \bibinfo {author}
  {\bibfnamefont {K.}~\bibnamefont {Watanabe}}, \bibinfo {author}
  {\bibfnamefont {B.~J.}\ \bibnamefont {LeRoy}}, \bibinfo {author}
  {\bibfnamefont {A.~H.}\ \bibnamefont {MacDonald}}, \ and\ \bibinfo {author}
  {\bibfnamefont {E.}~\bibnamefont {Tutuc}},\ }\href
  {http://www.pnas.org/content/early/2017/03/13/1620140114} {\bibinfo
  {journal} {PNAS, in press(2017)}\ }\BibitemShut {NoStop}%
\bibitem [{\citenamefont {Splendiani}\ \emph {et~al.}(2010)\citenamefont
  {Splendiani}, \citenamefont {Sun}, \citenamefont {Zhang}, \citenamefont {Li},
  \citenamefont {Kim}, \citenamefont {Chim}, \citenamefont {Galli},\ and\
  \citenamefont {Wang}}]{splendiani2010}%
  \BibitemOpen
\bibfield  {journal} {  }\bibfield  {author} {\bibinfo {author} {\bibfnamefont
  {A.}~\bibnamefont {Splendiani}}, \bibinfo {author} {\bibfnamefont
  {L.}~\bibnamefont {Sun}}, \bibinfo {author} {\bibfnamefont {Y.}~\bibnamefont
  {Zhang}}, \bibinfo {author} {\bibfnamefont {T.}~\bibnamefont {Li}}, \bibinfo
  {author} {\bibfnamefont {J.}~\bibnamefont {Kim}}, \bibinfo {author}
  {\bibfnamefont {C.-Y.}\ \bibnamefont {Chim}}, \bibinfo {author}
  {\bibfnamefont {G.}~\bibnamefont {Galli}}, \ and\ \bibinfo {author}
  {\bibfnamefont {F.}~\bibnamefont {Wang}},\ }\href
  {http://pubs.acs.org/doi/abs/10.1021/nl903868w} {\bibfield  {journal}
  {\bibinfo  {journal} {Nano Lett.}\ }\textbf {\bibinfo {volume} {10}},\
  \bibinfo {pages} {1271} (\bibinfo {year} {2010})}\BibitemShut {NoStop}%
\bibitem [{\citenamefont {Mak}\ \emph {et~al.}(2010)\citenamefont {Mak},
  \citenamefont {Lee}, \citenamefont {Hone}, \citenamefont {Shan},\ and\
  \citenamefont {Heinz}}]{mak2010}%
  \BibitemOpen
  \bibfield  {author} {\bibinfo {author} {\bibfnamefont {K.~F.}\ \bibnamefont
  {Mak}}, \bibinfo {author} {\bibfnamefont {C.}~\bibnamefont {Lee}}, \bibinfo
  {author} {\bibfnamefont {J.}~\bibnamefont {Hone}}, \bibinfo {author}
  {\bibfnamefont {J.}~\bibnamefont {Shan}}, \ and\ \bibinfo {author}
  {\bibfnamefont {T.~F.}\ \bibnamefont {Heinz}},\ }\href {\doibase
  10.1103/PhysRevLett.105.136805} {\bibfield  {journal} {\bibinfo  {journal}
  {Phys. Rev. Lett.}\ }\textbf {\bibinfo {volume} {105}},\ \bibinfo {pages}
  {136805} (\bibinfo {year} {2010})}\BibitemShut {NoStop}%
\bibitem [{\citenamefont {Ye}\ \emph {et~al.}(2014)\citenamefont {Ye},
  \citenamefont {Cao}, \citenamefont {O’Brien}, \citenamefont {Zhu},
  \citenamefont {Yin}, \citenamefont {Wang}, \citenamefont {Louie},\ and\
  \citenamefont {Zhang}}]{ye2014}%
  \BibitemOpen
  \bibfield  {author} {\bibinfo {author} {\bibfnamefont {Z.}~\bibnamefont
  {Ye}}, \bibinfo {author} {\bibfnamefont {T.}~\bibnamefont {Cao}}, \bibinfo
  {author} {\bibfnamefont {K.}~\bibnamefont {O’Brien}}, \bibinfo {author}
  {\bibfnamefont {H.}~\bibnamefont {Zhu}}, \bibinfo {author} {\bibfnamefont
  {X.}~\bibnamefont {Yin}}, \bibinfo {author} {\bibfnamefont {Y.}~\bibnamefont
  {Wang}}, \bibinfo {author} {\bibfnamefont {S.~G.}\ \bibnamefont {Louie}}, \
  and\ \bibinfo {author} {\bibfnamefont {X.}~\bibnamefont {Zhang}},\ }\href
  {http://www.nature.com/nature/journal/v513/n7517/full/nature13734.html}
  {\bibfield  {journal} {\bibinfo  {journal} {Nature}\ }\textbf {\bibinfo
  {volume} {513}},\ \bibinfo {pages} {214} (\bibinfo {year}
  {2014})}\BibitemShut {NoStop}%
\bibitem [{\citenamefont {He}\ \emph {et~al.}(2014)\citenamefont {He},
  \citenamefont {Kumar}, \citenamefont {Zhao}, \citenamefont {Wang},
  \citenamefont {Mak}, \citenamefont {Zhao},\ and\ \citenamefont
  {Shan}}]{He2014}%
  \BibitemOpen
  \bibfield  {author} {\bibinfo {author} {\bibfnamefont {K.}~\bibnamefont
  {He}}, \bibinfo {author} {\bibfnamefont {N.}~\bibnamefont {Kumar}}, \bibinfo
  {author} {\bibfnamefont {L.}~\bibnamefont {Zhao}}, \bibinfo {author}
  {\bibfnamefont {Z.}~\bibnamefont {Wang}}, \bibinfo {author} {\bibfnamefont
  {K.~F.}\ \bibnamefont {Mak}}, \bibinfo {author} {\bibfnamefont
  {H.}~\bibnamefont {Zhao}}, \ and\ \bibinfo {author} {\bibfnamefont
  {J.}~\bibnamefont {Shan}},\ }\href {\doibase 10.1103/PhysRevLett.113.026803}
  {\bibfield  {journal} {\bibinfo  {journal} {Phys. Rev. Lett.}\ }\textbf
  {\bibinfo {volume} {113}},\ \bibinfo {pages} {026803} (\bibinfo {year}
  {2014})}\BibitemShut {NoStop}%
\bibitem [{\citenamefont {Chernikov}\ \emph {et~al.}(2014)\citenamefont
  {Chernikov}, \citenamefont {Berkelbach}, \citenamefont {Hill}, \citenamefont
  {Rigosi}, \citenamefont {Li}, \citenamefont {Aslan}, \citenamefont
  {Reichman}, \citenamefont {Hybertsen},\ and\ \citenamefont
  {Heinz}}]{chernikov2014}%
  \BibitemOpen
  \bibfield  {author} {\bibinfo {author} {\bibfnamefont {A.}~\bibnamefont
  {Chernikov}}, \bibinfo {author} {\bibfnamefont {T.~C.}\ \bibnamefont
  {Berkelbach}}, \bibinfo {author} {\bibfnamefont {H.~M.}\ \bibnamefont
  {Hill}}, \bibinfo {author} {\bibfnamefont {A.}~\bibnamefont {Rigosi}},
  \bibinfo {author} {\bibfnamefont {Y.}~\bibnamefont {Li}}, \bibinfo {author}
  {\bibfnamefont {O.~B.}\ \bibnamefont {Aslan}}, \bibinfo {author}
  {\bibfnamefont {D.~R.}\ \bibnamefont {Reichman}}, \bibinfo {author}
  {\bibfnamefont {M.~S.}\ \bibnamefont {Hybertsen}}, \ and\ \bibinfo {author}
  {\bibfnamefont {T.~F.}\ \bibnamefont {Heinz}},\ }\href {\doibase
  10.1103/PhysRevLett.113.076802} {\bibfield  {journal} {\bibinfo  {journal}
  {Phys. Rev. Lett.}\ }\textbf {\bibinfo {volume} {113}},\ \bibinfo {pages}
  {076802} (\bibinfo {year} {2014})}\BibitemShut {NoStop}%
\bibitem [{\citenamefont {Qiu}\ \emph {et~al.}(2013)\citenamefont {Qiu},
  \citenamefont {da~Jornada},\ and\ \citenamefont {Louie}}]{Qiu2013}%
  \BibitemOpen
  \bibfield  {author} {\bibinfo {author} {\bibfnamefont {D.~Y.}\ \bibnamefont
  {Qiu}}, \bibinfo {author} {\bibfnamefont {F.~H.}\ \bibnamefont {da~Jornada}},
  \ and\ \bibinfo {author} {\bibfnamefont {S.~G.}\ \bibnamefont {Louie}},\
  }\href {\doibase 10.1103/PhysRevLett.111.216805} {\bibfield  {journal}
  {\bibinfo  {journal} {Phys. Rev. Lett.}\ }\textbf {\bibinfo {volume} {111}},\
  \bibinfo {pages} {216805} (\bibinfo {year} {2013})}\BibitemShut {NoStop}%
\bibitem [{\citenamefont {Cao}\ \emph {et~al.}(2012)\citenamefont {Cao},
  \citenamefont {Wang}, \citenamefont {Han}, \citenamefont {Ye}, \citenamefont
  {Zhu}, \citenamefont {Shi}, \citenamefont {Niu}, \citenamefont {Tan},
  \citenamefont {Wang}, \citenamefont {Liu} \emph {et~al.}}]{cao2012valley}%
  \BibitemOpen
  \bibfield  {author} {\bibinfo {author} {\bibfnamefont {T.}~\bibnamefont
  {Cao}}, \bibinfo {author} {\bibfnamefont {G.}~\bibnamefont {Wang}}, \bibinfo
  {author} {\bibfnamefont {W.}~\bibnamefont {Han}}, \bibinfo {author}
  {\bibfnamefont {H.}~\bibnamefont {Ye}}, \bibinfo {author} {\bibfnamefont
  {C.}~\bibnamefont {Zhu}}, \bibinfo {author} {\bibfnamefont {J.}~\bibnamefont
  {Shi}}, \bibinfo {author} {\bibfnamefont {Q.}~\bibnamefont {Niu}}, \bibinfo
  {author} {\bibfnamefont {P.}~\bibnamefont {Tan}}, \bibinfo {author}
  {\bibfnamefont {E.}~\bibnamefont {Wang}}, \bibinfo {author} {\bibfnamefont
  {B.}~\bibnamefont {Liu}},  \emph {et~al.},\ }\href
  {http://www.nature.com/articles/ncomms1882} {\bibfield  {journal} {\bibinfo
  {journal} {Nat. Commun.}\ }\textbf {\bibinfo {volume} {3}},\ \bibinfo {pages}
  {887} (\bibinfo {year} {2012})}\BibitemShut {NoStop}%
\bibitem [{\citenamefont {Zeng}\ \emph {et~al.}(2012)\citenamefont {Zeng},
  \citenamefont {Dai}, \citenamefont {Yao}, \citenamefont {Xiao},\ and\
  \citenamefont {Cui}}]{zeng2012valley}%
  \BibitemOpen
  \bibfield  {author} {\bibinfo {author} {\bibfnamefont {H.}~\bibnamefont
  {Zeng}}, \bibinfo {author} {\bibfnamefont {J.}~\bibnamefont {Dai}}, \bibinfo
  {author} {\bibfnamefont {W.}~\bibnamefont {Yao}}, \bibinfo {author}
  {\bibfnamefont {D.}~\bibnamefont {Xiao}}, \ and\ \bibinfo {author}
  {\bibfnamefont {X.}~\bibnamefont {Cui}},\ }\href
  {http://www.nature.com/nnano/journal/v7/n8/full/nnano.2012.95.html}
  {\bibfield  {journal} {\bibinfo  {journal} {Nat. Nanotechnol.}\ }\textbf
  {\bibinfo {volume} {7}},\ \bibinfo {pages} {490} (\bibinfo {year}
  {2012})}\BibitemShut {NoStop}%
\bibitem [{\citenamefont {Mak}\ \emph {et~al.}(2012)\citenamefont {Mak},
  \citenamefont {He}, \citenamefont {Shan},\ and\ \citenamefont
  {Heinz}}]{mak2012control}%
  \BibitemOpen
  \bibfield  {author} {\bibinfo {author} {\bibfnamefont {K.~F.}\ \bibnamefont
  {Mak}}, \bibinfo {author} {\bibfnamefont {K.}~\bibnamefont {He}}, \bibinfo
  {author} {\bibfnamefont {J.}~\bibnamefont {Shan}}, \ and\ \bibinfo {author}
  {\bibfnamefont {T.~F.}\ \bibnamefont {Heinz}},\ }\href
  {http://www.nature.com/nnano/journal/v7/n8/abs/nnano.2012.96.html} {\bibfield
   {journal} {\bibinfo  {journal} {Nat. Nanotechnol.}\ }\textbf {\bibinfo
  {volume} {7}},\ \bibinfo {pages} {494} (\bibinfo {year} {2012})}\BibitemShut
  {NoStop}%
\bibitem [{\citenamefont {Xiao}\ \emph {et~al.}(2012)\citenamefont {Xiao},
  \citenamefont {Liu}, \citenamefont {Feng}, \citenamefont {Xu},\ and\
  \citenamefont {Yao}}]{Di2012}%
  \BibitemOpen
  \bibfield  {author} {\bibinfo {author} {\bibfnamefont {D.}~\bibnamefont
  {Xiao}}, \bibinfo {author} {\bibfnamefont {G.-B.}\ \bibnamefont {Liu}},
  \bibinfo {author} {\bibfnamefont {W.}~\bibnamefont {Feng}}, \bibinfo {author}
  {\bibfnamefont {X.}~\bibnamefont {Xu}}, \ and\ \bibinfo {author}
  {\bibfnamefont {W.}~\bibnamefont {Yao}},\ }\href {\doibase
  10.1103/PhysRevLett.108.196802} {\bibfield  {journal} {\bibinfo  {journal}
  {Phys. Rev. Lett.}\ }\textbf {\bibinfo {volume} {108}},\ \bibinfo {pages}
  {196802} (\bibinfo {year} {2012})}\BibitemShut {NoStop}%
\bibitem [{\citenamefont {Mak}\ \emph {et~al.}(2014)\citenamefont {Mak},
  \citenamefont {McGill}, \citenamefont {Park},\ and\ \citenamefont
  {McEuen}}]{mak2014valley}%
  \BibitemOpen
  \bibfield  {author} {\bibinfo {author} {\bibfnamefont {K.~F.}\ \bibnamefont
  {Mak}}, \bibinfo {author} {\bibfnamefont {K.~L.}\ \bibnamefont {McGill}},
  \bibinfo {author} {\bibfnamefont {J.}~\bibnamefont {Park}}, \ and\ \bibinfo
  {author} {\bibfnamefont {P.~L.}\ \bibnamefont {McEuen}},\ }\href
  {http://science.sciencemag.org/content/344/6191/1489} {\bibfield  {journal}
  {\bibinfo  {journal} {Science}\ }\textbf {\bibinfo {volume} {344}},\ \bibinfo
  {pages} {1489} (\bibinfo {year} {2014})}\BibitemShut {NoStop}%
\bibitem [{\citenamefont {Kim}\ \emph {et~al.}(2014)\citenamefont {Kim},
  \citenamefont {Hong}, \citenamefont {Jin}, \citenamefont {Shi}, \citenamefont
  {Chang}, \citenamefont {Chiu}, \citenamefont {Li},\ and\ \citenamefont
  {Wang}}]{kim2014}%
  \BibitemOpen
  \bibfield  {author} {\bibinfo {author} {\bibfnamefont {J.}~\bibnamefont
  {Kim}}, \bibinfo {author} {\bibfnamefont {X.}~\bibnamefont {Hong}}, \bibinfo
  {author} {\bibfnamefont {C.}~\bibnamefont {Jin}}, \bibinfo {author}
  {\bibfnamefont {S.-F.}\ \bibnamefont {Shi}}, \bibinfo {author} {\bibfnamefont
  {C.-Y.~S.}\ \bibnamefont {Chang}}, \bibinfo {author} {\bibfnamefont {M.-H.}\
  \bibnamefont {Chiu}}, \bibinfo {author} {\bibfnamefont {L.-J.}\ \bibnamefont
  {Li}}, \ and\ \bibinfo {author} {\bibfnamefont {F.}~\bibnamefont {Wang}},\
  }\href {http://science.sciencemag.org/content/346/6214/1205} {\bibfield
  {journal} {\bibinfo  {journal} {Science}\ }\textbf {\bibinfo {volume}
  {346}},\ \bibinfo {pages} {1205} (\bibinfo {year} {2014})}\BibitemShut
  {NoStop}%
\bibitem [{\citenamefont {Sie}\ \emph {et~al.}(2015)\citenamefont {Sie},
  \citenamefont {McIver}, \citenamefont {Lee}, \citenamefont {Fu},
  \citenamefont {Kong},\ and\ \citenamefont {Gedik}}]{sie2015}%
  \BibitemOpen
  \bibfield  {author} {\bibinfo {author} {\bibfnamefont {E.~J.}\ \bibnamefont
  {Sie}}, \bibinfo {author} {\bibfnamefont {J.~W.}\ \bibnamefont {McIver}},
  \bibinfo {author} {\bibfnamefont {Y.-H.}\ \bibnamefont {Lee}}, \bibinfo
  {author} {\bibfnamefont {L.}~\bibnamefont {Fu}}, \bibinfo {author}
  {\bibfnamefont {J.}~\bibnamefont {Kong}}, \ and\ \bibinfo {author}
  {\bibfnamefont {N.}~\bibnamefont {Gedik}},\ }\href
  {http://www.nature.com/nmat/journal/v14/n3/abs/nmat4156.html} {\bibfield
  {journal} {\bibinfo  {journal} {Nat. Mat.}\ }\textbf {\bibinfo {volume}
  {14}},\ \bibinfo {pages} {290} (\bibinfo {year} {2015})}\BibitemShut
  {NoStop}%
\bibitem [{\citenamefont {Fang}\ \emph {et~al.}(2014)\citenamefont {Fang} \emph
  {et~al.}}]{Fang2014}%
  \BibitemOpen
  \bibfield  {author} {\bibinfo {author} {\bibfnamefont {H.}~\bibnamefont
  {Fang}} \emph {et~al.},\ }\href
  {http://www.pnas.org/content/111/17/6198.abstract} {\bibfield  {journal}
  {\bibinfo  {journal} {PNAS}\ }\textbf {\bibinfo {volume} {111}},\ \bibinfo
  {pages} {6198} (\bibinfo {year} {2014})}\BibitemShut {NoStop}%
\bibitem [{\citenamefont {Gong}\ \emph {et~al.}(2014)\citenamefont {Gong} \emph
  {et~al.}}]{gong2014}%
  \BibitemOpen
  \bibfield  {author} {\bibinfo {author} {\bibfnamefont {Y.}~\bibnamefont
  {Gong}} \emph {et~al.},\ }\href
  {http://www.nature.com/nmat/journal/v13/n12/full/nmat4091.html} {\bibfield
  {journal} {\bibinfo  {journal} {Nat. Mat.}\ }\textbf {\bibinfo {volume}
  {13}},\ \bibinfo {pages} {1135} (\bibinfo {year} {2014})}\BibitemShut
  {NoStop}%
\bibitem [{\citenamefont {Liu}\ \emph {et~al.}(2014)\citenamefont {Liu},
  \citenamefont {Zhang}, \citenamefont {Cao}, \citenamefont {Jin},
  \citenamefont {Qiu}, \citenamefont {Zhou}, \citenamefont {Zettl},
  \citenamefont {Yang}, \citenamefont {Louie},\ and\ \citenamefont
  {Wang}}]{liu2014evolution}%
  \BibitemOpen
  \bibfield  {author} {\bibinfo {author} {\bibfnamefont {K.}~\bibnamefont
  {Liu}}, \bibinfo {author} {\bibfnamefont {L.}~\bibnamefont {Zhang}}, \bibinfo
  {author} {\bibfnamefont {T.}~\bibnamefont {Cao}}, \bibinfo {author}
  {\bibfnamefont {C.}~\bibnamefont {Jin}}, \bibinfo {author} {\bibfnamefont
  {D.}~\bibnamefont {Qiu}}, \bibinfo {author} {\bibfnamefont {Q.}~\bibnamefont
  {Zhou}}, \bibinfo {author} {\bibfnamefont {A.}~\bibnamefont {Zettl}},
  \bibinfo {author} {\bibfnamefont {P.}~\bibnamefont {Yang}}, \bibinfo {author}
  {\bibfnamefont {S.~G.}\ \bibnamefont {Louie}}, \ and\ \bibinfo {author}
  {\bibfnamefont {F.}~\bibnamefont {Wang}},\ }\href
  {http://www.nature.com/articles/ncomms5966?WT.ec_id=NCOMMS-20140924}
  {\bibfield  {journal} {\bibinfo  {journal} {Nat. Commun.}\ }\textbf {\bibinfo
  {volume} {5}},\ \bibinfo {pages} {4966} (\bibinfo {year} {2014})}\BibitemShut
  {NoStop}%
\bibitem [{\citenamefont {Rivera}\ \emph {et~al.}(2016)\citenamefont {Rivera},
  \citenamefont {Seyler}, \citenamefont {Yu}, \citenamefont {Schaibley},
  \citenamefont {Yan}, \citenamefont {Mandrus}, \citenamefont {Yao},\ and\
  \citenamefont {Xu}}]{rivera2016valley}%
  \BibitemOpen
  \bibfield  {author} {\bibinfo {author} {\bibfnamefont {P.}~\bibnamefont
  {Rivera}}, \bibinfo {author} {\bibfnamefont {K.~L.}\ \bibnamefont {Seyler}},
  \bibinfo {author} {\bibfnamefont {H.}~\bibnamefont {Yu}}, \bibinfo {author}
  {\bibfnamefont {J.~R.}\ \bibnamefont {Schaibley}}, \bibinfo {author}
  {\bibfnamefont {J.}~\bibnamefont {Yan}}, \bibinfo {author} {\bibfnamefont
  {D.~G.}\ \bibnamefont {Mandrus}}, \bibinfo {author} {\bibfnamefont
  {W.}~\bibnamefont {Yao}}, \ and\ \bibinfo {author} {\bibfnamefont
  {X.}~\bibnamefont {Xu}},\ }\href
  {http://science.sciencemag.org/content/351/6274/688} {\bibfield  {journal}
  {\bibinfo  {journal} {Science}\ }\textbf {\bibinfo {volume} {351}},\ \bibinfo
  {pages} {688} (\bibinfo {year} {2016})}\BibitemShut {NoStop}%
\bibitem [{\citenamefont {Yu}\ \emph {et~al.}(2015)\citenamefont {Yu},
  \citenamefont {Wang}, \citenamefont {Tong}, \citenamefont {Xu},\ and\
  \citenamefont {Yao}}]{Yu2015}%
  \BibitemOpen
  \bibfield  {author} {\bibinfo {author} {\bibfnamefont {H.}~\bibnamefont
  {Yu}}, \bibinfo {author} {\bibfnamefont {Y.}~\bibnamefont {Wang}}, \bibinfo
  {author} {\bibfnamefont {Q.}~\bibnamefont {Tong}}, \bibinfo {author}
  {\bibfnamefont {X.}~\bibnamefont {Xu}}, \ and\ \bibinfo {author}
  {\bibfnamefont {W.}~\bibnamefont {Yao}},\ }\href {\doibase
  10.1103/PhysRevLett.115.187002} {\bibfield  {journal} {\bibinfo  {journal}
  {Phys. Rev. Lett.}\ }\textbf {\bibinfo {volume} {115}},\ \bibinfo {pages}
  {187002} (\bibinfo {year} {2015})}\BibitemShut {NoStop}%
\bibitem [{\citenamefont {Fogler}\ \emph {et~al.}(2014)\citenamefont {Fogler},
  \citenamefont {Butov},\ and\ \citenamefont {Novoselov}}]{fogler2014high}%
  \BibitemOpen
  \bibfield  {author} {\bibinfo {author} {\bibfnamefont {M.}~\bibnamefont
  {Fogler}}, \bibinfo {author} {\bibfnamefont {L.}~\bibnamefont {Butov}}, \
  and\ \bibinfo {author} {\bibfnamefont {K.}~\bibnamefont {Novoselov}},\ }\href
  {http://www.nature.com/articles/ncomms5555} {\bibfield  {journal} {\bibinfo
  {journal} {Nat. Commun.}\ }\textbf {\bibinfo {volume} {5}} (\bibinfo {year}
  {2014})}\BibitemShut {NoStop}%
\bibitem [{\citenamefont {Wu}\ \emph {et~al.}(2015{\natexlab{a}})\citenamefont
  {Wu}, \citenamefont {Xue},\ and\ \citenamefont {MacDonald}}]{Wu2015}%
  \BibitemOpen
  \bibfield  {author} {\bibinfo {author} {\bibfnamefont {F.-C.}\ \bibnamefont
  {Wu}}, \bibinfo {author} {\bibfnamefont {F.}~\bibnamefont {Xue}}, \ and\
  \bibinfo {author} {\bibfnamefont {A.~H.}\ \bibnamefont {MacDonald}},\ }\href
  {\doibase 10.1103/PhysRevB.92.165121} {\bibfield  {journal} {\bibinfo
  {journal} {Phys. Rev. B}\ }\textbf {\bibinfo {volume} {92}},\ \bibinfo
  {pages} {165121} (\bibinfo {year} {2015}{\natexlab{a}})}\BibitemShut
  {NoStop}%
\bibitem [{\citenamefont {Yu}\ \emph {et~al.}(2014)\citenamefont {Yu},
  \citenamefont {Liu}, \citenamefont {Gong}, \citenamefont {Xu},\ and\
  \citenamefont {Yao}}]{yu2014dirac}%
  \BibitemOpen
  \bibfield  {author} {\bibinfo {author} {\bibfnamefont {H.}~\bibnamefont
  {Yu}}, \bibinfo {author} {\bibfnamefont {G.-B.}\ \bibnamefont {Liu}},
  \bibinfo {author} {\bibfnamefont {P.}~\bibnamefont {Gong}}, \bibinfo {author}
  {\bibfnamefont {X.}~\bibnamefont {Xu}}, \ and\ \bibinfo {author}
  {\bibfnamefont {W.}~\bibnamefont {Yao}},\ }\href
  {http://www.nature.com/articles/ncomms4876?WT.ec_id=NCOMMS-20140514}
  {\bibfield  {journal} {\bibinfo  {journal} {Nat. Commun.}\ }\textbf {\bibinfo
  {volume} {5}} (\bibinfo {year} {2014})}\BibitemShut {NoStop}%
\bibitem [{\citenamefont {Glazov}\ \emph {et~al.}(2014)\citenamefont {Glazov},
  \citenamefont {Amand}, \citenamefont {Marie}, \citenamefont {Lagarde},
  \citenamefont {Bouet},\ and\ \citenamefont {Urbaszek}}]{Glazov2014}%
  \BibitemOpen
  \bibfield  {author} {\bibinfo {author} {\bibfnamefont {M.~M.}\ \bibnamefont
  {Glazov}}, \bibinfo {author} {\bibfnamefont {T.}~\bibnamefont {Amand}},
  \bibinfo {author} {\bibfnamefont {X.}~\bibnamefont {Marie}}, \bibinfo
  {author} {\bibfnamefont {D.}~\bibnamefont {Lagarde}}, \bibinfo {author}
  {\bibfnamefont {L.}~\bibnamefont {Bouet}}, \ and\ \bibinfo {author}
  {\bibfnamefont {B.}~\bibnamefont {Urbaszek}},\ }\href {\doibase
  10.1103/PhysRevB.89.201302} {\bibfield  {journal} {\bibinfo  {journal} {Phys.
  Rev. B}\ }\textbf {\bibinfo {volume} {89}},\ \bibinfo {pages} {201302}
  (\bibinfo {year} {2014})}\BibitemShut {NoStop}%
\bibitem [{\citenamefont {Yu}\ and\ \citenamefont {Wu}(2014)}]{Yu2014}%
  \BibitemOpen
  \bibfield  {author} {\bibinfo {author} {\bibfnamefont {T.}~\bibnamefont
  {Yu}}\ and\ \bibinfo {author} {\bibfnamefont {M.~W.}\ \bibnamefont {Wu}},\
  }\href {\doibase 10.1103/PhysRevB.89.205303} {\bibfield  {journal} {\bibinfo
  {journal} {Phys. Rev. B}\ }\textbf {\bibinfo {volume} {89}},\ \bibinfo
  {pages} {205303} (\bibinfo {year} {2014})}\BibitemShut {NoStop}%
\bibitem [{\citenamefont {Wu}\ \emph {et~al.}(2015{\natexlab{b}})\citenamefont
  {Wu}, \citenamefont {Qu},\ and\ \citenamefont {MacDonald}}]{wu2015Exciton}%
  \BibitemOpen
  \bibfield  {author} {\bibinfo {author} {\bibfnamefont {F.}~\bibnamefont
  {Wu}}, \bibinfo {author} {\bibfnamefont {F.}~\bibnamefont {Qu}}, \ and\
  \bibinfo {author} {\bibfnamefont {A.~H.}\ \bibnamefont {MacDonald}},\ }\href
  {\doibase 10.1103/PhysRevB.91.075310} {\bibfield  {journal} {\bibinfo
  {journal} {Phys. Rev. B}\ }\textbf {\bibinfo {volume} {91}},\ \bibinfo
  {pages} {075310} (\bibinfo {year} {2015}{\natexlab{b}})}\BibitemShut
  {NoStop}%
\bibitem [{Note1()}]{Note1}%
  \BibitemOpen
  \bibinfo {note} {The effect of the exchange interaction on spatially indirect
  excitons has been discussed, for example, in Ref.~\cite
  {Durnev2016}}\BibitemShut {NoStop}%
\bibitem [{\citenamefont {Yuen-Zhou}\ \emph {et~al.}(2014)\citenamefont
  {Yuen-Zhou}, \citenamefont {Saikin}, \citenamefont {Yao},\ and\ \citenamefont
  {Aspuru-Guzik}}]{yuen2014topologically}%
  \BibitemOpen
  \bibfield  {author} {\bibinfo {author} {\bibfnamefont {J.}~\bibnamefont
  {Yuen-Zhou}}, \bibinfo {author} {\bibfnamefont {S.~K.}\ \bibnamefont
  {Saikin}}, \bibinfo {author} {\bibfnamefont {N.~Y.}\ \bibnamefont {Yao}}, \
  and\ \bibinfo {author} {\bibfnamefont {A.}~\bibnamefont {Aspuru-Guzik}},\
  }\href {http://www.nature.com/nmat/journal/v13/n11/abs/nmat4073.html}
  {\bibfield  {journal} {\bibinfo  {journal} {Nat. Mat.}\ }\textbf {\bibinfo
  {volume} {13}},\ \bibinfo {pages} {1026} (\bibinfo {year}
  {2014})}\BibitemShut {NoStop}%
\bibitem [{\citenamefont {Karzig}\ \emph {et~al.}(2015)\citenamefont {Karzig},
  \citenamefont {Bardyn}, \citenamefont {Lindner},\ and\ \citenamefont
  {Refael}}]{Karzig2015}%
  \BibitemOpen
  \bibfield  {author} {\bibinfo {author} {\bibfnamefont {T.}~\bibnamefont
  {Karzig}}, \bibinfo {author} {\bibfnamefont {C.-E.}\ \bibnamefont {Bardyn}},
  \bibinfo {author} {\bibfnamefont {N.~H.}\ \bibnamefont {Lindner}}, \ and\
  \bibinfo {author} {\bibfnamefont {G.}~\bibnamefont {Refael}},\ }\href
  {\doibase 10.1103/PhysRevX.5.031001} {\bibfield  {journal} {\bibinfo
  {journal} {Phys. Rev. X}\ }\textbf {\bibinfo {volume} {5}},\ \bibinfo {pages}
  {031001} (\bibinfo {year} {2015})}\BibitemShut {NoStop}%
\bibitem [{\citenamefont {Bardyn}\ \emph {et~al.}(2015)\citenamefont {Bardyn},
  \citenamefont {Karzig}, \citenamefont {Refael},\ and\ \citenamefont
  {Liew}}]{Bardyn2015}%
  \BibitemOpen
  \bibfield  {author} {\bibinfo {author} {\bibfnamefont {C.-E.}\ \bibnamefont
  {Bardyn}}, \bibinfo {author} {\bibfnamefont {T.}~\bibnamefont {Karzig}},
  \bibinfo {author} {\bibfnamefont {G.}~\bibnamefont {Refael}}, \ and\ \bibinfo
  {author} {\bibfnamefont {T.~C.~H.}\ \bibnamefont {Liew}},\ }\href {\doibase
  10.1103/PhysRevB.91.161413} {\bibfield  {journal} {\bibinfo  {journal} {Phys.
  Rev. B}\ }\textbf {\bibinfo {volume} {91}},\ \bibinfo {pages} {161413}
  (\bibinfo {year} {2015})}\BibitemShut {NoStop}%
\bibitem [{\citenamefont {Nalitov}\ \emph {et~al.}(2015)\citenamefont
  {Nalitov}, \citenamefont {Solnyshkov},\ and\ \citenamefont
  {Malpuech}}]{Nalitov2015}%
  \BibitemOpen
  \bibfield  {author} {\bibinfo {author} {\bibfnamefont {A.~V.}\ \bibnamefont
  {Nalitov}}, \bibinfo {author} {\bibfnamefont {D.~D.}\ \bibnamefont
  {Solnyshkov}}, \ and\ \bibinfo {author} {\bibfnamefont {G.}~\bibnamefont
  {Malpuech}},\ }\href {\doibase 10.1103/PhysRevLett.114.116401} {\bibfield
  {journal} {\bibinfo  {journal} {Phys. Rev. Lett.}\ }\textbf {\bibinfo
  {volume} {114}},\ \bibinfo {pages} {116401} (\bibinfo {year}
  {2015})}\BibitemShut {NoStop}%
\bibitem [{\citenamefont {Song}\ and\ \citenamefont {Rudner}(2016)}]{Song2016}%
  \BibitemOpen
  \bibfield  {author} {\bibinfo {author} {\bibfnamefont {J.~C.~W.}\
  \bibnamefont {Song}}\ and\ \bibinfo {author} {\bibfnamefont {M.~S.}\
  \bibnamefont {Rudner}},\ }\href {\doibase 10.1073/pnas.1519086113} {\bibfield
   {journal} {\bibinfo  {journal} {PNAS}\ }\textbf {\bibinfo {volume} {113}},\
  \bibinfo {pages} {4658} (\bibinfo {year} {2016})}\BibitemShut {NoStop}%
\bibitem [{\citenamefont {Jin}\ \emph {et~al.}(2016)\citenamefont {Jin},
  \citenamefont {Lu}, \citenamefont {Wang}, \citenamefont {Fang}, \citenamefont
  {Joannopoulos}, \citenamefont {Solja{\v{c}}i{\'c}}, \citenamefont {Fu},\ and\
  \citenamefont {Fang}}]{jin2016}%
  \BibitemOpen
  \bibfield  {author} {\bibinfo {author} {\bibfnamefont {D.}~\bibnamefont
  {Jin}}, \bibinfo {author} {\bibfnamefont {L.}~\bibnamefont {Lu}}, \bibinfo
  {author} {\bibfnamefont {Z.}~\bibnamefont {Wang}}, \bibinfo {author}
  {\bibfnamefont {C.}~\bibnamefont {Fang}}, \bibinfo {author} {\bibfnamefont
  {J.~D.}\ \bibnamefont {Joannopoulos}}, \bibinfo {author} {\bibfnamefont
  {M.}~\bibnamefont {Solja{\v{c}}i{\'c}}}, \bibinfo {author} {\bibfnamefont
  {L.}~\bibnamefont {Fu}}, \ and\ \bibinfo {author} {\bibfnamefont {N.~X.}\
  \bibnamefont {Fang}},\ }\href {http://www.nature.com/articles/ncomms13486}
  {\bibfield  {journal} {\bibinfo  {journal} {Nat. Commun.}\ }\textbf {\bibinfo
  {volume} {7}},\ \bibinfo {pages} {13486} (\bibinfo {year}
  {2016})}\BibitemShut {NoStop}%
\bibitem [{\citenamefont {Wilson}\ \emph {et~al.}(2017)\citenamefont {Wilson}
  \emph {et~al.}}]{wilson2016band}%
  \BibitemOpen
  \bibfield  {author} {\bibinfo {author} {\bibfnamefont {N.~R.}\ \bibnamefont
  {Wilson}} \emph {et~al.},\ }\href
  {http://advances.sciencemag.org/content/3/2/e1601832} {\bibfield  {journal}
  {\bibinfo  {journal} {Sci. Adv.}\ }\textbf {\bibinfo {volume} {3}},\ \bibinfo
  {pages} {e1601832} (\bibinfo {year} {2017})}\BibitemShut {NoStop}%
\bibitem [{\citenamefont {Hamann}(2013)}]{hamann2013}%
  \BibitemOpen
  \bibfield  {author} {\bibinfo {author} {\bibfnamefont {D.~R.}\ \bibnamefont
  {Hamann}},\ }\href {\doibase 10.1103/PhysRevB.88.085117} {\bibfield
  {journal} {\bibinfo  {journal} {Phys. Rev. B}\ }\textbf {\bibinfo {volume}
  {88}},\ \bibinfo {pages} {085117} (\bibinfo {year} {2013})}\BibitemShut
  {NoStop}%
\bibitem [{\citenamefont {Schlipf}\ and\ \citenamefont
  {Gygi}(2015)}]{schlipf2015}%
  \BibitemOpen
  \bibfield  {author} {\bibinfo {author} {\bibfnamefont {M.}~\bibnamefont
  {Schlipf}}\ and\ \bibinfo {author} {\bibfnamefont {F.}~\bibnamefont {Gygi}},\
  }\href {http://www.sciencedirect.com/science/article/pii/S0010465515001897}
  {\bibfield  {journal} {\bibinfo  {journal} {Comput. Phys. Commun.}\ }\textbf
  {\bibinfo {volume} {196}},\ \bibinfo {pages} {36} (\bibinfo {year}
  {2015})}\BibitemShut {NoStop}%
\bibitem [{\citenamefont {Giannozzi}\ \emph {et~al.}(2009)\citenamefont
  {Giannozzi} \emph {et~al.}}]{giannozzi2009}%
  \BibitemOpen
  \bibfield  {author} {\bibinfo {author} {\bibfnamefont {P.}~\bibnamefont
  {Giannozzi}} \emph {et~al.},\ }\href
  {http://iopscience.iop.org/article/10.1088/0953-8984/21/39/395502/meta}
  {\bibfield  {journal} {\bibinfo  {journal} {J. Phys.: Condens. Matter}\
  }\textbf {\bibinfo {volume} {21}},\ \bibinfo {pages} {395502} (\bibinfo
  {year} {2009})}\BibitemShut {NoStop}%
\bibitem [{\citenamefont {Mostofi}\ \emph {et~al.}(2014)\citenamefont
  {Mostofi}, \citenamefont {Yates}, \citenamefont {Pizzi}, \citenamefont {Lee},
  \citenamefont {Souza}, \citenamefont {Vanderbilt},\ and\ \citenamefont
  {Marzari}}]{mostofi2014}%
  \BibitemOpen
  \bibfield  {author} {\bibinfo {author} {\bibfnamefont {A.~A.}\ \bibnamefont
  {Mostofi}}, \bibinfo {author} {\bibfnamefont {J.~R.}\ \bibnamefont {Yates}},
  \bibinfo {author} {\bibfnamefont {G.}~\bibnamefont {Pizzi}}, \bibinfo
  {author} {\bibfnamefont {Y.-S.}\ \bibnamefont {Lee}}, \bibinfo {author}
  {\bibfnamefont {I.}~\bibnamefont {Souza}}, \bibinfo {author} {\bibfnamefont
  {D.}~\bibnamefont {Vanderbilt}}, \ and\ \bibinfo {author} {\bibfnamefont
  {N.}~\bibnamefont {Marzari}},\ }\href {\doibase
  http://dx.doi.org/10.1016/j.cpc.2014.05.003} {\bibfield  {journal} {\bibinfo
  {journal} {Comput. Phys. Commun.}\ }\textbf {\bibinfo {volume} {185}},\
  \bibinfo {pages} {2309} (\bibinfo {year} {2014})}\BibitemShut {NoStop}%
\bibitem [{SM()}]{SM}%
  \BibitemOpen
  \href@noop {} {\bibinfo  {journal} {See Supplemental Material for details of
  {\em ab initio} calculations, discussion on local approximation, topological
  phase diagram and edge state analysis. It includes Refs.~\cite{perdew1981,
  rasmussen2015, yun2012, dalcorso2014}}\ }\BibitemShut {NoStop}%
\bibitem [{Note2()}]{Note2}%
  \BibitemOpen
\bibfield  {journal} {  }\bibinfo {note} {Bulk properties of moir\'e systems
  with finite twist angle are independent of the relative displacement prior to
  twist which we set to zero. See Ref.~\protect \rev@citealp {Bistritzer2011}
  for an explanation.}\BibitemShut {Stop}%
\bibitem [{\citenamefont {Wu}\ \emph {et~al.}(2014)\citenamefont {Wu},
  \citenamefont {Qian},\ and\ \citenamefont {Li}}]{wu2014tunable}%
  \BibitemOpen
  \bibfield  {author} {\bibinfo {author} {\bibfnamefont {M.}~\bibnamefont
  {Wu}}, \bibinfo {author} {\bibfnamefont {X.}~\bibnamefont {Qian}}, \ and\
  \bibinfo {author} {\bibfnamefont {J.}~\bibnamefont {Li}},\ }\href
  {http://pubs.acs.org/doi/abs/10.1021/nl502414t} {\bibfield  {journal}
  {\bibinfo  {journal} {Nano Lett.}\ }\textbf {\bibinfo {volume} {14}},\
  \bibinfo {pages} {5350} (\bibinfo {year} {2014})}\BibitemShut {NoStop}%
\bibitem [{\citenamefont {Jung}\ \emph {et~al.}(2014)\citenamefont {Jung},
  \citenamefont {Raoux}, \citenamefont {Qiao},\ and\ \citenamefont
  {MacDonald}}]{Jung2014}%
  \BibitemOpen
  \bibfield  {author} {\bibinfo {author} {\bibfnamefont {J.}~\bibnamefont
  {Jung}}, \bibinfo {author} {\bibfnamefont {A.}~\bibnamefont {Raoux}},
  \bibinfo {author} {\bibfnamefont {Z.}~\bibnamefont {Qiao}}, \ and\ \bibinfo
  {author} {\bibfnamefont {A.~H.}\ \bibnamefont {MacDonald}},\ }\href {\doibase
  10.1103/PhysRevB.89.205414} {\bibfield  {journal} {\bibinfo  {journal} {Phys.
  Rev. B}\ }\textbf {\bibinfo {volume} {89}},\ \bibinfo {pages} {205414}
  (\bibinfo {year} {2014})}\BibitemShut {NoStop}%
\bibitem [{Note3()}]{Note3}%
  \BibitemOpen
  \bibinfo {note} {We used static approximation for the exchange interaction.
  Retardation effect results in intrinsic energy broadening of exciton states
  in the light cone. See Ref.~\cite {Glazov2014}.}\BibitemShut {Stop}%
\bibitem [{\citenamefont {Qiu}\ \emph {et~al.}(2015)\citenamefont {Qiu},
  \citenamefont {Cao},\ and\ \citenamefont {Louie}}]{Qiu2015}%
  \BibitemOpen
  \bibfield  {author} {\bibinfo {author} {\bibfnamefont {D.~Y.}\ \bibnamefont
  {Qiu}}, \bibinfo {author} {\bibfnamefont {T.}~\bibnamefont {Cao}}, \ and\
  \bibinfo {author} {\bibfnamefont {S.~G.}\ \bibnamefont {Louie}},\ }\href
  {\doibase 10.1103/PhysRevLett.115.176801} {\bibfield  {journal} {\bibinfo
  {journal} {Phys. Rev. Lett.}\ }\textbf {\bibinfo {volume} {115}},\ \bibinfo
  {pages} {176801} (\bibinfo {year} {2015})}\BibitemShut {NoStop}%
\bibitem [{\citenamefont {MacNeill}\ \emph {et~al.}(2015)\citenamefont
  {MacNeill}, \citenamefont {Heikes}, \citenamefont {Mak}, \citenamefont
  {Anderson}, \citenamefont {Korm\'anyos}, \citenamefont {Z\'olyomi},
  \citenamefont {Park},\ and\ \citenamefont {Ralph}}]{MacNeil2015}%
  \BibitemOpen
  \bibfield  {author} {\bibinfo {author} {\bibfnamefont {D.}~\bibnamefont
  {MacNeill}}, \bibinfo {author} {\bibfnamefont {C.}~\bibnamefont {Heikes}},
  \bibinfo {author} {\bibfnamefont {K.~F.}\ \bibnamefont {Mak}}, \bibinfo
  {author} {\bibfnamefont {Z.}~\bibnamefont {Anderson}}, \bibinfo {author}
  {\bibfnamefont {A.}~\bibnamefont {Korm\'anyos}}, \bibinfo {author}
  {\bibfnamefont {V.}~\bibnamefont {Z\'olyomi}}, \bibinfo {author}
  {\bibfnamefont {J.}~\bibnamefont {Park}}, \ and\ \bibinfo {author}
  {\bibfnamefont {D.~C.}\ \bibnamefont {Ralph}},\ }\href {\doibase
  10.1103/PhysRevLett.114.037401} {\bibfield  {journal} {\bibinfo  {journal}
  {Phys. Rev. Lett.}\ }\textbf {\bibinfo {volume} {114}},\ \bibinfo {pages}
  {037401} (\bibinfo {year} {2015})}\BibitemShut {NoStop}%
\bibitem [{\citenamefont {Srivastava}\ \emph {et~al.}(2015)\citenamefont
  {Srivastava}, \citenamefont {Sidler}, \citenamefont {Allain}, \citenamefont
  {Lembke}, \citenamefont {Kis},\ and\ \citenamefont
  {Imamo{\u{g}}lu}}]{srivastava2015valley}%
  \BibitemOpen
  \bibfield  {author} {\bibinfo {author} {\bibfnamefont {A.}~\bibnamefont
  {Srivastava}}, \bibinfo {author} {\bibfnamefont {M.}~\bibnamefont {Sidler}},
  \bibinfo {author} {\bibfnamefont {A.~V.}\ \bibnamefont {Allain}}, \bibinfo
  {author} {\bibfnamefont {D.~S.}\ \bibnamefont {Lembke}}, \bibinfo {author}
  {\bibfnamefont {A.}~\bibnamefont {Kis}}, \ and\ \bibinfo {author}
  {\bibfnamefont {A.}~\bibnamefont {Imamo{\u{g}}lu}},\ }\href
  {http://www.nature.com/nphys/journal/vaop/ncurrent/full/nphys3203.html}
  {\bibfield  {journal} {\bibinfo  {journal} {Nat. Phys.}\ }\textbf {\bibinfo
  {volume} {11}},\ \bibinfo {pages} {141} (\bibinfo {year} {2015})}\BibitemShut
  {NoStop}%
\bibitem [{\citenamefont {Aivazian}\ \emph {et~al.}(2015)\citenamefont
  {Aivazian}, \citenamefont {Gong}, \citenamefont {Jones}, \citenamefont {Chu},
  \citenamefont {Yan}, \citenamefont {Mandrus}, \citenamefont {Zhang},
  \citenamefont {Cobden}, \citenamefont {Yao},\ and\ \citenamefont
  {Xu}}]{aivazian2015magnetic}%
  \BibitemOpen
  \bibfield  {author} {\bibinfo {author} {\bibfnamefont {G.}~\bibnamefont
  {Aivazian}}, \bibinfo {author} {\bibfnamefont {Z.}~\bibnamefont {Gong}},
  \bibinfo {author} {\bibfnamefont {A.~M.}\ \bibnamefont {Jones}}, \bibinfo
  {author} {\bibfnamefont {R.-L.}\ \bibnamefont {Chu}}, \bibinfo {author}
  {\bibfnamefont {J.}~\bibnamefont {Yan}}, \bibinfo {author} {\bibfnamefont
  {D.~G.}\ \bibnamefont {Mandrus}}, \bibinfo {author} {\bibfnamefont
  {C.}~\bibnamefont {Zhang}}, \bibinfo {author} {\bibfnamefont
  {D.}~\bibnamefont {Cobden}}, \bibinfo {author} {\bibfnamefont
  {W.}~\bibnamefont {Yao}}, \ and\ \bibinfo {author} {\bibfnamefont
  {X.}~\bibnamefont {Xu}},\ }\href
  {http://www.nature.com/nphys/journal/v11/n2/full/nphys3201.html} {\bibfield
  {journal} {\bibinfo  {journal} {Nat. Phys.}\ }\textbf {\bibinfo {volume}
  {11}},\ \bibinfo {pages} {148} (\bibinfo {year} {2015})}\BibitemShut
  {NoStop}%
\bibitem [{\citenamefont {Durnev}\ and\ \citenamefont
  {Glazov}(2016)}]{Durnev2016}%
  \BibitemOpen
  \bibfield  {author} {\bibinfo {author} {\bibfnamefont {M.~V.}\ \bibnamefont
  {Durnev}}\ and\ \bibinfo {author} {\bibfnamefont {M.~M.}\ \bibnamefont
  {Glazov}},\ }\href {\doibase 10.1103/PhysRevB.93.155409} {\bibfield
  {journal} {\bibinfo  {journal} {Phys. Rev. B}\ }\textbf {\bibinfo {volume}
  {93}},\ \bibinfo {pages} {155409} (\bibinfo {year} {2016})}\BibitemShut
  {NoStop}%
\bibitem [{\citenamefont {Perdew}\ and\ \citenamefont
  {Zunger}(1981)}]{perdew1981}%
  \BibitemOpen
  \bibfield  {author} {\bibinfo {author} {\bibfnamefont {J.~P.}\ \bibnamefont
  {Perdew}}\ and\ \bibinfo {author} {\bibfnamefont {A.}~\bibnamefont
  {Zunger}},\ }\href {\doibase 10.1103/PhysRevB.23.5048} {\bibfield  {journal}
  {\bibinfo  {journal} {Phys. Rev. B}\ }\textbf {\bibinfo {volume} {23}},\
  \bibinfo {pages} {5048} (\bibinfo {year} {1981})}\BibitemShut {NoStop}%
\bibitem [{\citenamefont {Rasmussen}\ and\ \citenamefont
  {Thygesen}(2015)}]{rasmussen2015}%
  \BibitemOpen
  \bibfield  {author} {\bibinfo {author} {\bibfnamefont {F.~A.}\ \bibnamefont
  {Rasmussen}}\ and\ \bibinfo {author} {\bibfnamefont {K.~S.}\ \bibnamefont
  {Thygesen}},\ }\href {http://pubs.acs.org/doi/abs/10.1021/acs.jpcc.5b02950}
  {\bibfield  {journal} {\bibinfo  {journal} {J. Phys. Chem. C}\ }\textbf
  {\bibinfo {volume} {119}},\ \bibinfo {pages} {13169} (\bibinfo {year}
  {2015})}\BibitemShut {NoStop}%
\bibitem [{\citenamefont {Yun}\ \emph {et~al.}(2012)\citenamefont {Yun},
  \citenamefont {Han}, \citenamefont {Hong}, \citenamefont {Kim},\ and\
  \citenamefont {Lee}}]{yun2012}%
  \BibitemOpen
  \bibfield  {author} {\bibinfo {author} {\bibfnamefont {W.~S.}\ \bibnamefont
  {Yun}}, \bibinfo {author} {\bibfnamefont {S.~W.}\ \bibnamefont {Han}},
  \bibinfo {author} {\bibfnamefont {S.~C.}\ \bibnamefont {Hong}}, \bibinfo
  {author} {\bibfnamefont {I.~G.}\ \bibnamefont {Kim}}, \ and\ \bibinfo
  {author} {\bibfnamefont {J.~D.}\ \bibnamefont {Lee}},\ }\href {\doibase
  10.1103/PhysRevB.85.033305} {\bibfield  {journal} {\bibinfo  {journal} {Phys.
  Rev. B}\ }\textbf {\bibinfo {volume} {85}},\ \bibinfo {pages} {033305}
  (\bibinfo {year} {2012})}\BibitemShut {NoStop}%
\bibitem [{\citenamefont {Corso}(2014)}]{dalcorso2014}%
  \BibitemOpen
  \bibfield  {author} {\bibinfo {author} {\bibfnamefont {A.~D.}\ \bibnamefont
  {Corso}},\ }\href
  {http://www.sciencedirect.com/science/article/pii/S0927025614005187}
  {\bibfield  {journal} {\bibinfo  {journal} {Computational Materials Science}\
  }\textbf {\bibinfo {volume} {95}},\ \bibinfo {pages} {337} (\bibinfo {year}
  {2014})}\BibitemShut {NoStop}%
\bibitem [{\citenamefont {Bistritzer}\ and\ \citenamefont
  {MacDonald}(2011)}]{Bistritzer2011}%
  \BibitemOpen
  \bibfield  {author} {\bibinfo {author} {\bibfnamefont {R.}~\bibnamefont
  {Bistritzer}}\ and\ \bibinfo {author} {\bibfnamefont {A.~H.}\ \bibnamefont
  {MacDonald}},\ }\href {\doibase 10.1073/pnas.1108174108} {\bibfield
  {journal} {\bibinfo  {journal} {PNAS}\ }\textbf {\bibinfo {volume} {108}},\
  \bibinfo {pages} {12233} (\bibinfo {year} {2011})}\BibitemShut {NoStop}%
\end{thebibliography}%

\clearpage
\begin{center}
\textbf{Supplemental Material}
\end{center}

\section{Details of the {\em ab initio} calculation}
We perform fully-relativistic density functional theory (DFT) calculations under the local-density approximation (LDA) \cite{perdew1981} for the MoS$_2$/WS$_2$ bilayer system using the Quantum Espresso distribution \cite{giannozzi2009}.
We use norm-conserving pseudopotentials based on those provided by the SG15 pseudopotential library \cite{schlipf2015}; this library provides optimized inputs for the ONCVPSP pseudopotential generation method \cite{hamann2013}.
We alter the pseudopotential generation inputs from the SG15 library to produce fully-relativistic, LDA pseudopotentials; the cutoff radius of the W $p$ states is also reduced slightly to eliminate a ghost state which appears when converting the pseudopotential to fully-relativistic LDA.

Lattice structures for the MoS$_2$ and WS$_2$ layers are taken from Ref.~\cite{rasmussen2015}.
We take the interlayer distance to be that of bulk MoS$_2$ as given by Ref.~\cite{yun2012} and choose a 20\AA vacuum distance for the slab model.
We choose a plane-wave cutoff energy of 60 Ry and corresponding charge density cutoff energy of 240 Ry, and we sample the Brillouin zone with a 9$\times$9$\times$1 grid of $k$ points.
The gap values obtained using this cutoff energy and $k$-space grid are found to be well-converged relative to calculations at twice the cutoff energy and at twice the number of $k$ points in each direction.
The variation of the overall band gap with $\boldsymbol{d}$ is also found to be consistent with a calculation in which we replaced the norm-conserving pseudopotentials with projector augmented wave (PAW) data sets from PSlibrary 1.0.0 \cite{dalcorso2014}.

To determine the orbital character of electronic bands, we obtain a tight-binding Hamiltonian in a basis of Wannier functions using Wannier90 \cite{mostofi2014}. We project onto $p$ orbitals for S and $d$ orbitals for Mo and W, and reproduce the DFT band structure with high accuracy. Disentanglement is necessary due to a small overlap in energy between the group of conduction bands and higher states; we do not perform maximal localization. The gap between valence and conduction bands for MoS$_2$ states at the $K$ point (shown in Fig.~1(b,e) of the main text) is calculated by considering states with total weight greater than a threshold value of 0.7 on the MoS$_2$ layer.
We find that the states nearest the Fermi energy at the $K$ point are strongly layer-selective, with the gap value obtained by this procedure being independent of this threshold in the range 0.05 to 0.85 for both AA and AB stacking.

The software used to generate inputs for Quantum Espresso and Wannier90 and to calculate gaps in MoS$_2$-dominated states from the {\em ab initio} result is available at \url{https://github.com/tflovorn/tmd}.

%Previous work shows that LDA provides a reasonable description of bilayer graphene and graphite when compared to more sophisticated and computationally
%intensive approaches, including quantum Monte Carlo and the random phase approximation \cite{gould2013, rego2015}.
It is known that LDA underestimates the band gap for semiconductors like MoS$_2$.
We view our work as a qualitative instead of quantitative study, and expect that the effects discussed in the main text are generic.
Moreover our strategy for estimating the exciton moir\'e potential strength can accommodate more accurate approaches, like GW calculation and GW Bethe-Salpeter calculation.  

\section{local approximation for exciton moir\'e potential}
In this section we discuss the local approximation we use for the influence 
of substrate registration on the exciton Hamiltonian.
There are three length scales in the problem: the lattice constant $a_0$, the exciton size $a_X$ and the moir\'e periodicity $a_M$.
For MoS$_2$, $a_0$ is about $3.19{\text \AA}$. 
The moir\'e length $a_M = a_0/\theta$ is about $20$nm for a typical 
twist angle $\theta$ around $1^{\circ}$.
We can approximate $a_X$ by the root of mean square of the electron-hole separation in 
the $A$ exciton state of MoS$_2$.
The exciton wave function is obtained by solving the Bethe-Salpeter equation based on massive Dirac model. The details of this microscopic model can be found in Appendix B of Ref.\cite{wu2015Exciton}. $a_X$ depends on dielectric constant $\varepsilon$. From that calculation, we find that $a_X$ is 1.05nm, 1.15nm and 1.3nm respectively for $\varepsilon=$1, 2.5 and 4. Note that monolayer MoS$_2$ is two dimensional, and its dielectric constant is mainly determined by its environment. For example $\varepsilon$ is about 4 when the twisted bilayer is encapsulated by hexagonal boron nitride. 
As $\varepsilon$ increases, the Coulomb interaction becomes weaker and meantime the
local screening effect from the 2D material itself
also becomes weaker\cite{wu2015Exciton}. This is the reason why $a_X$ does not scale linearly with $\varepsilon$.

Because $a_X>a_0$, a $k\cdot p$ description of exciton is justified. 
As $a_M$ is much larger than $a_X$ and the band gap varies smoothly on the scale of $a_M$, it is a good approximation to assume that the variation of exciton energy is determined by the local band gap and band edge mass parameters.

The exciton energy is $E_g-E_b$, where $E_g$ and $E_b$ are respectively band gap and exciton binding energy. 
We have obtained the variation in $E_g$ from {\em ab initio} calculation as reported in the main text.  
We now provide an estimate of the spatial variation of $E_b$.
In $k\cdot p$ theory $E_b$ scales linearly with the reduced mass $\mu=m_c m_v/(m_c+m_v)$, where $m_c$ and $m_v$ are respectively the conduction and valence band effective mass. $\mu$ has been calculated as a function of the relative displacement $\boldsymbol{d}$ based on DFT band structure. The numerical results show that $\mu$ has a variation about $\pm 0.5\%$, which leads to a similar relative variation in $E_b$. It is known that $E_b$ in monolayer TMDs is relatively large and on the order of few hundred meV\cite{Qiu2013,ye2014,He2014,chernikov2014,wu2015Exciton}. The exact value of $E_b$ depends on the dielectric constant $\varepsilon$. If we estimate $E_b$ as 400meV, then the variation of $E_b$ due to moir\'e pattern is about $\pm 2$meV. Note that the variation of $E_g$ is about $\pm 10$meV, as shown in Fig. 1(b) and (e) in the main text. From this estimation, we argue that the variation of $E_g$ dominates over that of $E_b$.  With this justification we neglect the variation in $E_b$. 

We emphasize that the parametrization of $\Delta(\boldsymbol{r})$, i.e. the exciton energy variation in moir\'e pattern, in Eq.(4) of the main text results from the three-fold rotational symmetry of the lattice structure and the smoothness of moir\'e potential.  The parametrization of $\Delta(\boldsymbol{r})$ in terms of two parameters $(V, \psi)$ is valid 
independent of the microscopic origin of the exciton energy variation.

\section{Topological phase diagram}

\begin{figure}[t]
	\includegraphics[width=1\columnwidth]{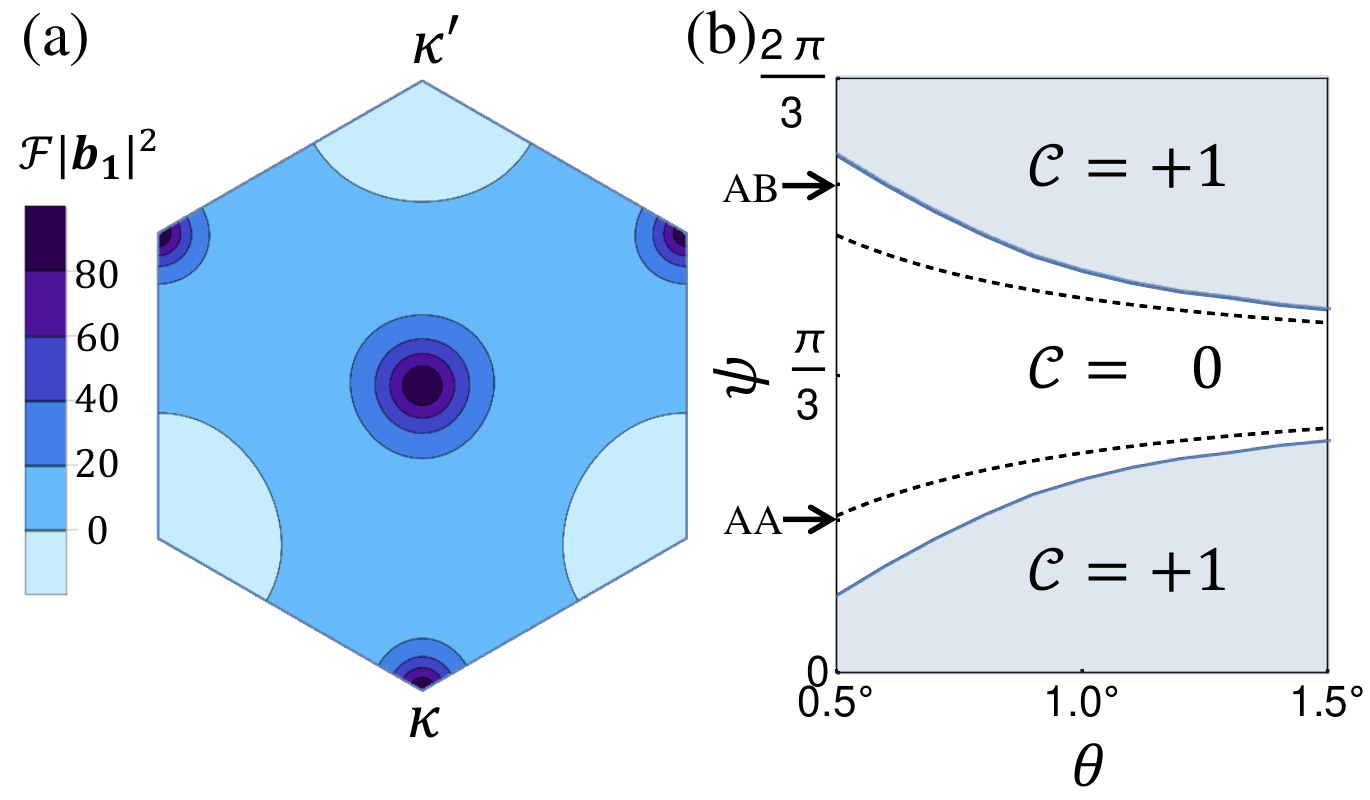}
	\caption{(color online). 	(a) Berry curvature $\mathcal{F}$ of the first band in Fig. 3(a) of the main text. $\mathcal{F}$ is calculated using the Kubo formula expression.
	(b) Chern number $\mathcal{C}$ of the first exciton moir\'e band as a function of $\theta$ and $\psi$ with ($V$, $h_z$) fixed at
	 (1.4meV, 1.5meV). The arrows indicate the value of $\psi$ in AA and AB stacked MoS$_2$/WS$_2$. Adding $2\pi/3$ to $\psi$ leads 
	 to a spatial translation of the moir\'e pattern. It follows that $\mathcal{C}$ is a periodic function of $\psi$ with periodicity $2\pi/3$. Dashed lines mark analytic but approximate phase boundaries determined by (\ref{pkappa}) and (\ref{pkappap}). 
	 }
	\label{Berry_PD}
\end{figure}

The Berry curvature $\mathcal{F}$ of the first exciton band in Fig. 3(b) of the main text is shown in Fig.~\ref{Berry_PD}(a), which displays hot spots around $\gamma$, $\kappa$, and $\kappa'$ points in the MBZ.  
$\mathcal{F}$ around $\gamma$ is simply understood 
in terms of the valley Berry phase induced by the exchange interaction, and its sign is determined by that of $h_z$. 
The peak in $\mathcal{F}$ around the $\kappa$ and $\kappa'$ point is related to gap opening in moir\'e pattern, 
and can be sensitive to $\psi$, the phase of $V_1$. 
To test how robust the topological exciton band is, we calculate the Chern number of the first band as a function of $\psi$ and twist angle $\theta$.
The resulting topological phase diagram is shown in Fig.~\ref{Berry_PD}(b).
Because the Chern number is finite in a large parameter space of $(\psi, \theta)$, 
we expect that there is a good chance that topological exciton bands can be realized in TMD bilayers. 
Along the phase boundary lines at which the Chern number changes, 
the energy gaps at $\kappa$ or $\kappa'$ close.  
Below we show that the phase boundaries can be qualitatively captured by treating the moir\'e potential as a perturbation. 

In the absence of the moir\'e pattern, there are two energy modes at each momentum. The wave function of the lower mode at momentum $\boldsymbol{Q}$ can be expressed as a two-component spinor in the valley space:
\begin{equation}
|\boldsymbol{Q}\rangle = \begin{pmatrix} \sin(\frac{\alpha_{\boldsymbol{Q}}}{2}) \\ -\cos(\frac{\alpha_{\boldsymbol{Q}}}{2}) e^{i 2 \phi_{\boldsymbol{Q}}} \end{pmatrix},
\end{equation}
where $\alpha_{\boldsymbol{Q}}$ is determined by the condition:
\begin{equation}
\cos(\alpha_{\boldsymbol{Q}}) = \frac{h_z}{\sqrt{h_z^2+(J |\boldsymbol{Q}|)^2}}.
\end{equation}

\begin{figure}[tb]
	\includegraphics[width=1\columnwidth]{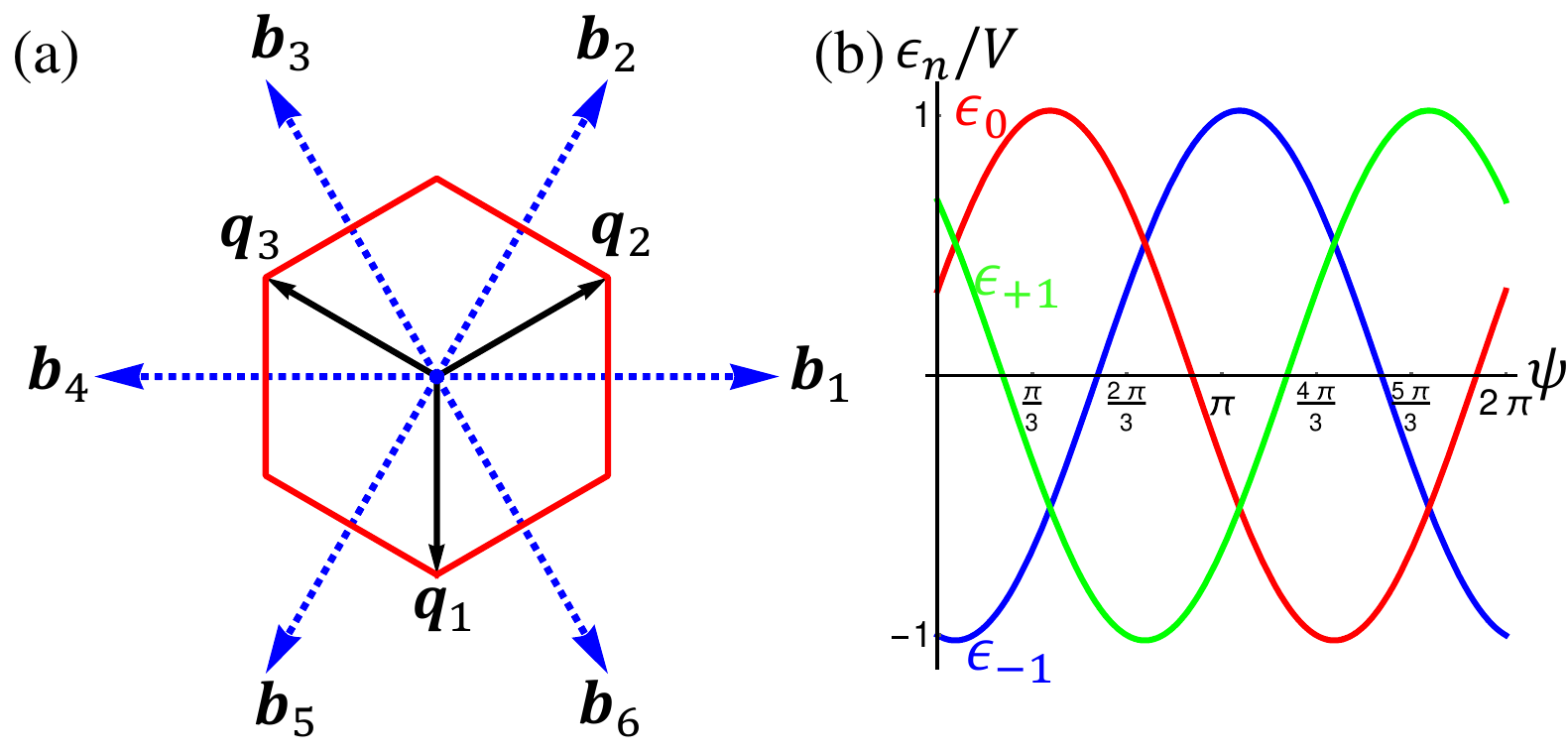}
	\caption{(color online). (a)Due to the moir\'e potential, states at three different momenta $\boldsymbol{q}_i$ are folded to the same point $\kappa$. (b) $\epsilon_n$ as a function $\psi$ at $\beta=0.2$.
	 }
	\label{analytic}
\end{figure}

The moir\'e potential couples states with momenta that differ by a moir\'e reciprocal lattice vector. At the $\kappa$ point of the moir\'e Brillouine zone, the three lowest energy states  mainly originate from $|\boldsymbol{q}_i\rangle$, the lower energy modes at momenta $\boldsymbol{q}_i$, which are illustrated in Fig. \ref{analytic}(a). The three lowest energy states at $\kappa$ would be degenerate if the moir\'e potential vanishes. The potential lifts the degeneracy, and can be treated by degenerate state perturbation theory. The potential projected to $|\boldsymbol{q}_i\rangle$ has the matrix form:
\begin{equation}
\begin{aligned}
\mathcal{V}_\kappa=&\begin{pmatrix}
0                                            & V_5 \langle \boldsymbol{q}_1 | \boldsymbol{q}_2 \rangle  & V_6 \langle \boldsymbol{q}_1 | \boldsymbol{q}_3 \rangle\\
V_2 \langle \boldsymbol{q}_2 | \boldsymbol{q}_1 \rangle  &                                            0 & V_1 \langle \boldsymbol{q}_2 | \boldsymbol{q}_3 \rangle \\
V_3 \langle \boldsymbol{q}_3 | \boldsymbol{q}_1 \rangle &V_4 \langle \boldsymbol{q}_3 | \boldsymbol{q}_2 \rangle    & 0  
\end{pmatrix} \\
=&\frac{V}{2}\begin{pmatrix}
0          & \lambda  & \lambda^* \\
\lambda^*  &        0 & \lambda \\
\lambda    & \lambda^*& 0  
\end{pmatrix},
\end{aligned}
\end{equation}
where parameter $\lambda$ is:
\begin{equation}
\lambda= e^{i\psi}[e^{-i\pi/3}-\beta e^{i \pi/6}], \, \beta= \frac{\sqrt{3} h_z}{\sqrt{h_z^2+(J |\boldsymbol{q}_1|)^2}}.
\label{lambdabeta}
\end{equation}
Note that $\psi$ is the phase of $V_1$, and $|\boldsymbol{q}_1| = 4\pi/(3 a_M)=4\theta \pi/(3 a_0)$. Therefore $|\boldsymbol{q}_1|$ is controlled by the twist angle $\theta$.

$\kappa$ is a high symmetry point. In particular, it is invariant under a three-fold rotational symmetry. Due to this symmetry, $\mathcal{V}_\kappa$ commutes with the following cyclic permutation matrix $\hat{C}_3$:
\begin{equation}
\hat{C}_3=\begin{pmatrix}
0 & 1 & 0\\
0 & 0 & 1\\
1 & 0 & 0
\end{pmatrix}.
\end{equation}

Thus $\mathcal{V}_\kappa$ and $\hat{C}_3$ can be diagonalized simultaneously:
\begin{equation}
\begin{aligned}
\mathcal{V}_\kappa |n\rangle & = \epsilon_n |n\rangle,\\
\hat{C}_3 |n\rangle & = e^{i 2n\pi/3} |n\rangle,
\end{aligned}
\end{equation}
where $n$ takes integer values -1, 0 and +1. The eigen states and energies are respectively:
\begin{equation}
\begin{aligned}
|n\rangle&=\frac{1}{\sqrt{3}}\begin{pmatrix}
e^{i 2n\pi/3} \\ e^{-i 2n\pi/3} \\1
\end{pmatrix},\\
\epsilon_n/V &= -\sqrt{1+\beta^2}\cos[\psi+(n+1)\frac{2\pi}{3}-\arctan \beta].
\end{aligned}
\end{equation}

Fig. \ref{analytic}(b) plots $\epsilon_n$ as a function of $\psi$ for a non-zero $\beta$ [defined in Eq.~(\ref{lambdabeta})]. Because of the symmetry $\hat{C}_3$, there are level crossings (instead of avoided crossing) between different states. If we consider $\psi$ in the range $(0, 2\pi/3)$ and assume $h_z$ is small, the gap between the lowest energy state and higher energy states at $\kappa$ closes when $\epsilon_{-1}$ equals $\epsilon_{+1}$. This leads to the $\kappa$ point gap closing condition:
\begin{equation}
\psi=\pi/3+\arctan \beta.
\label{pkappa}
\end{equation}

In the same manner, we find the gap closing condition at $\kappa'$ point is:
\begin{equation}
\psi=\pi/3-\arctan \beta.
\label{pkappap}
\end{equation}

Note that when $h_z=0$, $\beta$ vanishes and the gap at $\kappa$ and $\kappa'$ close simultaneously, which is a result of time reversal symmetry.
The analytic phase boundaries presented by Eqs. (\ref{pkappa}) and (\ref{pkappap}) correctly capture the trend of the numerical phase boundaries, as shown in Fig.~\ref{Berry_PD}(b).

\section{Energy Spectrum of a Stripe}
\begin{figure*}[t]
	\includegraphics[width=1.4\columnwidth]{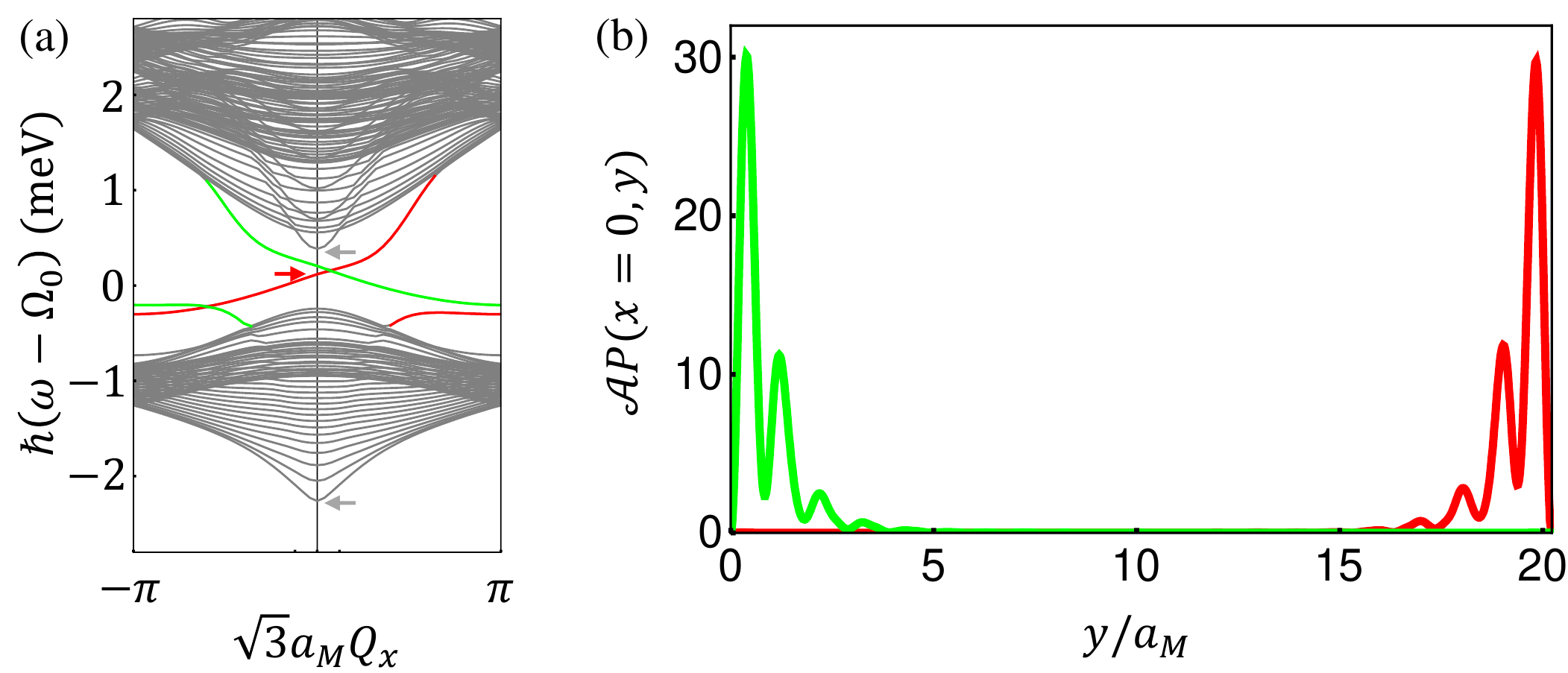}
	\caption{(color online). (a) Same figure as Fig. 3(b) in the main text. The stripe width $L_y$ is $20.22a_M$ and the vector $\boldsymbol{l}$ is taken to be (0,0.1)$a_M$. (b) Green (red) curve presents the spatial distribution of the $Q_x=0$ states on the green (red) line in (a).
	 }
	\label{Stripe}
\end{figure*}

To study edge states, we consider a stripe with finite width $L_y$ in $y$ direction.  We use a hard-wall boundary condition, which requires the wave function to vanish at the two edges, $y=0$ and $y=L_y$.  This boundary condition leads to the basis states:
\begin{equation}
|Q_x, n, \alpha \rangle = \sqrt{\frac{2}{\mathcal{A}}} \sin(\frac{n\pi}{L_y}y)\exp(i Q_x x)|\alpha\rangle,
\label{basis}
\end{equation}
where $\mathcal{A}$ is the area of the stripe and $n$ takes positive integer numbers. $|\alpha\rangle$ represents valley $K_+$ or $K_-$ exciton. 

We decompose the Hamiltonian into different terms:
\begin{equation}
\begin{aligned}
H&=\hat{h}_0+\hat{h}_x+\hat{h}_y+\hat{h}_z+ \sum_{j=1}^6 \tilde{V}_j \exp( i \boldsymbol{b}_j\cdot \boldsymbol{r})\tau_0,\\
\hat{h}_0&= (\hbar\Omega_0+\frac{\hbar^2 \boldsymbol{Q}^2}{2 M}+ J |\boldsymbol{Q}|)\tau_0,\\
\hat{h}_x&= J |\boldsymbol{Q}|\cos(2 \phi_{\boldsymbol{Q}})\tau_x ,\,\,
\hat{h}_y= J |\boldsymbol{Q}|\sin(2 \phi_{\boldsymbol{Q}})\tau_y,\\
\hat{h}_z&= h_z \tau_z, \,\,\,\,\,\,\,\,\,\,\,\,\,\,\,\,\,\,\,\,\,\,\,\,\,\,\,\,\,\,\,\,\,
\tilde{V}_j=V_j\exp(-i \boldsymbol{b}_j\cdot \boldsymbol{l}),
\end{aligned}
\label{Hstripe}
\end{equation}
where $\boldsymbol{Q}$ is understood to be an operator, and its eigenstates are plane waves. 
The vector $\boldsymbol{l}$ describes the spatial translation of the moir\'e pattern, which is not important for the bulk energy spectrum. 
However, it determines where the edge at $y=0$ is located in the moir\'e unit cell. 
$\boldsymbol{l}$ can be varied to tune the energy dispersion of states localized on the $y=0$ edge.

The Hamiltonian matrix element in the basis $|Q_x, n, \alpha \rangle$ is specified as follows:
\begin{widetext}
\begin{equation}
\begin{aligned}
\langle Q_x, n, \alpha | \hat{h}_0+\hat{h}_z |Q_x', n', \alpha' \rangle = & \delta_{Q_xQ_x'}\delta_{nn'}\delta_{\alpha\alpha'}[\hbar\Omega_0+\frac{\hbar^2 \boldsymbol{Q}^2}{2 M}+J |\boldsymbol{Q}|+h_z\tau_z^{(\alpha\alpha')}],\\
\langle Q_x, n, \alpha | \hat{h}_x |Q_x', n', \alpha' \rangle = & \delta_{Q_xQ_x'}\delta_{nn'} J |\boldsymbol{Q}|\cos(2 \phi_{\boldsymbol{Q}}) \tau_x^{(\alpha\alpha')},\\
\langle Q_x, n, \alpha | \hat{h}_y |Q_x', n', \alpha' \rangle = & \delta_{Q_xQ_x'} J \Big( \frac{Q_x}{\sqrt{(L_yQ_x)^2+(n\pi)^2}} + \frac{Q_x}{\sqrt{(L_yQ_x)^2+(n'\pi)^2}} \Big) f(n,n') \tau_y^{(\alpha\alpha')},\\
\langle Q_x, n, \alpha | \exp( i \boldsymbol{b}\cdot \boldsymbol{r})\tau_0 |Q_x', n', \alpha' \rangle = & \delta_{Q_x-Q_x', b_x}\delta_{\alpha\alpha'}\frac{1}{2}\Big[F(b_yL_y/\pi+n-n')+F(b_yL_y/\pi-n+n')\\&-F(b_yL_y/\pi+n+n')-F(b_yL_y/\pi-n-n')\Big],
\end{aligned}
\end{equation}
\end{widetext}
where $\boldsymbol{Q}$ represents the vector $(Q_x, \frac{n\pi}{L_y})$, and the two functions are:
\begin{equation}
\begin{aligned}
f(n,n')=&-i\pi n' \frac{2}{L_y}\int_0^{L_y} \sin(\frac{n \pi}{L_y}y)\cos(\frac{n' \pi}{L_y}y)dy\\
=&\begin{cases} 
      -i \frac{2 n n'}{n^2-n'^2}\big[1-(-1)^{n+n'}\big], & n\neq n' \\
      0, & n=n'
   \end{cases},
\end{aligned}   
\end{equation}

\begin{equation}
\begin{aligned}
F(z)=&\frac{1}{L_y}\int_0^{L_y} \exp(i \frac{z \pi}{L_y}y) dy \\
    =& \begin{cases} 
      i \frac{1-(-1)^z}{z\pi}, & z\neq 0 \\
      1, & z=0
   \end{cases}.
\end{aligned}
\end{equation}

The energy spectrum for the stripe geometry is obtained by diagonalizing the Hamiltonian matrix with a proper truncation in parameters $Q_x$ and $n$.
Because of the moir\'e potential, $Q_x$ reduced to the 1D Brillouin zone $(-\frac{\pi}{\sqrt{3}a_M},\frac{\pi}{\sqrt{3}a_M})$ is a good quantum number.
We define the spatial distribution of exciton state as:
\begin{equation}
P(\boldsymbol{r})=|\psi_{K_+}(\boldsymbol{r})|^2+|\psi_{K_-}(\boldsymbol{r})|^2,
\end{equation}
where $(\psi_{K_+}(\boldsymbol{r}), \psi_{K_-}(\boldsymbol{r}))$ represents valley spinor wave function of exciton.

Fig.~\ref{Stripe}(a) reproduces Fig.~3(b) in the main text. The red and blue curves highlight in-gap states, which are localized on opposite edges as demonstrated in Fig.~\ref{Stripe}(b). 
The localization length of edge states is governed by $a_M$, which is much bigger than the lattice constant.
The real edge on atomic length scale can be complicated. However, it should play a minor role on the exciton edge states because of different length scales.

\section{optical response of edge states}

%\begin{figure*}[t]
%	\includegraphics[width=1.4\columnwidth]{local_response.pdf}
%	\caption{(color online). Spatially resolved optical conductivity. (a) Response from bulk states. The energy splitting between the two peaks is due to %the Zeeman energy. (b) Response from both bulk and edge states. There is an enhancement of the optical response around the edge due to edge states. The %arrows indicate the energy level of bulk states (gray arrow) and edge state (red arrow) in the absence of energy broadening, and correspond to the three %arrows shown in Fig.~\ref{Stripe}(a).
%	 }
%	\label{local_response}
%\end{figure*}

Light-matter coupling is theoretically described by:
\begin{equation}
\delta H=-\int d \boldsymbol{r} \boldsymbol{A}\cdot\boldsymbol{j}(\boldsymbol{r}),
\end{equation}
where $\boldsymbol{j}$ is the current operator of matter, and $\boldsymbol{A}$ is the vector potential of light. 
We assume that light propagates perpendicular to the 2D system under study. Therefore, $\boldsymbol{A}$ is in the 2D plane, and is independent of the 2D position $\boldsymbol{r}$.

Using the Kubo formula, we can express the spatially resolved optical conductivity as follows:
\begin{equation}
\sigma_{ab}(\omega, \boldsymbol{r})=\frac{i}{\omega} \sum_{\chi} \frac{\langle G | j_a(\boldsymbol{r}) | \chi \rangle 
\langle \chi | \int d\boldsymbol{r}' j_b(\boldsymbol{r}') | G \rangle}{\hbar(\omega-\omega_\chi)+i \eta},
\label{localcond}
\end{equation}
where $|G\rangle$ and $|\chi\rangle$ respectively represent ground state and excited states.
Equation~(\ref{localcond}) describes the  optical response at position $\boldsymbol{r}$ to a spatially {\em uniform} electromagnetic field,
and $\omega$ is assumed to be at positive frequency.

It is instructive to consider the scaling of $\sigma(\boldsymbol{r})$ as a function of the spatial extension $W$ over which the exciton center-of-mass wave function $|\chi\rangle$ spreads. While the optical matrix element $\langle G | j_a(\boldsymbol{r}) | \chi \rangle$ scales as $1/\sqrt{W}$ because $|\chi\rangle$ is normalized, the spatial integration over $\boldsymbol{r}'$ in (\ref{localcond}) results in an extra factor $W$.  Therefore, the local optical conductivity $\sigma(\boldsymbol{r})$ is an {\it intensive} quantity with respect to $W$. This scaling analysis suggests that edge states with a small lateral spatial extension can have a local optical response comparable in magnitude to that of the bulk state. The explicit calculation below agrees with this reasoning.  

To study the optical response of a finite-width stripe, we make the following Fourier transformation:
\begin{equation}
\boldsymbol{j}(\boldsymbol{r}) = \frac{1}{\mathcal{A}}\sum_{Q_x, n} \sqrt{2}\sin(\frac{n\pi}{L_y}y) e^{-i Q_x x} \boldsymbol{j}(Q_x, n),
\end{equation}
where $y$ extends from 0 to $L_y$, and $n$ denotes positive integers.

We are interested in the variation of the optical response in $y$ direction due to the edge states. Thus, the optical conductivity is averaged over $x$ direction,
\begin{equation}
\begin{aligned}
&\sigma_{ab}(\omega, y)\\
=&\frac{1}{L_x}\int d x \sigma_{ab}(\omega, \boldsymbol{r})\\
=&\frac{i}{\omega\mathcal{A}}\sum_{\chi}\Big\{\Big[\sum_{n} \sqrt{2}\sin(\frac{n\pi}{L_y}y) \langle G | j_a(Q_x=0, n) |\chi \rangle \Big]\\
&\times  \Big[\sum_{n' \in \text{odd}} \frac{2\sqrt{2}}{n'\pi} \langle \chi | j_b(Q_x'=0, n') | G \rangle \Big]\\
&\times\frac{1}{\hbar(\omega-\omega_\chi)+i \eta}\Big\}.
\end{aligned}
\end{equation}

$|\chi\rangle$ can be expressed in terms of the basis states in Eq.~(\ref{basis}):
\begin{equation}
|\chi\rangle=\sum_{Q_x,n,\alpha}\chi_{Q_x,n,\alpha}|Q_x,n,\alpha\rangle,
\end{equation}
where $\chi_{Q_x,n,\alpha}$ represents the eigen vectors of the Hamiltonian in Eq.~(\ref{Hstripe}). 

To proceed, we make the following approximation:
\begin{equation}
\begin{aligned}
        &\langle Q_x, n, \alpha| \boldsymbol{j}(Q_x, n) |G\rangle \\ 
\approx & \langle Q_x=0, n=1, \alpha| \boldsymbol{j}(Q_x=0, n=1) |G\rangle,
\end{aligned}
\end{equation}
which is a small-momentum expansion of the optical matrix element to zeroth order.

Finally, we obtain the expression that is suitable for numerical calculation:
\begin{equation}
\begin{aligned}
\sigma_{xx}(\omega, y)&\approx \frac{\text{Re}\sigma_{xx}^{(0)}(\Omega_0)}{2}\sum_{\chi}\Big\{\Big[\sum_{n,\alpha} \sqrt{2}\sin(\frac{n\pi}{L_y}y) \chi_{0,n,\alpha}\Big]\\
&\times \Big[\sum_{n'\in \text{odd},\alpha'} \frac{2\sqrt{2}}{n' \pi} \chi_{0,n',\alpha'}^*\Big]
\frac{i \eta}{\hbar(\omega-\omega_\chi)+i \eta}\Big\},
\end{aligned}
\label{localsigmaxx}
\end{equation}
where $\text{Re}\sigma_{xx}^{(0)}(\Omega_0)$ is the $A$ exciton optical conductivity peak in the bulk and in the absence of Zeeman field.

For bulk states, we assume that their optical response is spatially uniform. Because of the Zeeman field, the $A$ exciton peak splits into two peaks around $\Omega_0$. The optical conductivity from bulk states is illustrated in Fig.~4(a) of the main text. The broadening factor $\eta$ is chosen to be 1meV.

The optical response due to edge states is calculated using Eq.~(\ref{localsigmaxx}). The overall optical conductivity including contribution from both bulk and edge states is shown in Fig.~4(b)  of the main text.  The edge states have a strong  optical response near the edge. Based on numerical results, we find that the {\it maximum} local optical conductivity due to one edge state is about $0.19\text{Re}\sigma_{xx}^{(0)}(\Omega_0)$, which is comparable in magnitude to that of the bulk optical response. Note that we have used a broadening factor $\eta$ (1meV) that exceeds the bulk exciton band gap (see Fig.~\ref{Stripe}(a)). Nevertheless, the enhancement of the optical response near the edge is clearly visible.

Exciton states have a finite lifetime due to exciton-light coupling, which leads to intrinsic energy broadening. In our study, this intrinsic broadening is phenomenologically captured by the parameter $\eta$.

\end{document}